\documentclass[manuscript,nonacm]{acmart}

\AtBeginDocument{%
  }

\usepackage{latexsym}

\usepackage[T1]{fontenc}

\usepackage[utf8]{inputenc}

\usepackage{graphicx}

\usepackage{todonotes}      %
\usepackage{booktabs}
\usepackage{amsmath}
\usepackage{listings}
\usepackage{subcaption}
\usepackage[title]{appendix}

\usepackage{float}

\setcopyright{none}

\setuptodonotes{disable}   %
\begin{document}

\title{WebQuest: A Benchmark for Multimodal QA on Web Page Sequences}

\author{Maria Wang}
\authornote{Joint first authors.}
\author{Srinivas Sunkara}
\authornotemark[1]
\author{Gilles Baechler}
\authornote{Joint second authors.}
\author{Jason Lin}
\authornotemark[2]
\author{Yun Zhu}
\author{Fedir Zubach}
\author{Lei Shu}
\author{Jindong Chen}
\affiliation{%
  \institution{Google DeepMind}
  \country{USA}
}

\authorsaddresses{%
Correspondence: jdchen@google.com
}

\renewcommand{\shortauthors}{Wang et al.}

\begin{abstract}

The rise of powerful multimodal LLMs has enhanced the viability of building web agents which can, with increasing levels of autonomy, assist users to retrieve information and complete tasks on various human-computer interfaces. It is hence necessary to build challenging benchmarks that span a wide-variety of use cases reflecting real-world usage. In this work, we present WebQuest, a multi-page question-answering dataset that requires reasoning across multiple related web pages. In contrast to existing UI benchmarks that focus on multi-step web navigation and task completion, our dataset evaluates information extraction, multimodal retrieval and composition of information from many web pages. WebQuest includes three question categories: single-screen QA, multi-screen QA, and QA based on navigation traces. We evaluate leading proprietary multimodal models like GPT-4V, Gemini Flash, Claude 3, and open source models like InstructBLIP, PaliGemma on our dataset, revealing a significant gap between  single-screen and multi-screen reasoning. Finally, we investigate inference time techniques like Chain-of-Thought prompting to improve model capabilities on multi-screen reasoning.

\end{abstract}


\maketitle

\section{Introduction}
\label{sec:intro}
As the de facto I/O for Human-Computer Interaction, Graphical User-Interfaces (UIs) act as a crucial bridge between computers and humans, allowing users to acquire information, complete tasks, engage in social interactions and perform various other daily activities. A predominant  type of data we consume are multimodal web pages and apps rich in semantic knowledge and structure. The ubiquity and utility of UIs has motivated research for multimodal, assistive technology to simplify various user journeys. At the same time, recent multimodal LLMs~\cite{geminiteam2024gemini,openai2024gpt4,anthropic2024claude3} with strong capabilities across a variety of tasks have sparked interest in building end-to-end agents adept at assisting users with various complicated tasks. 

One of the challenges in developing UI models is to interpret spatial and semantic relations among the various contents in sequences of screens. Traditional computer vision tasks like summarization, image captioning, visual question answering have been extended to document images~\cite{mathew2021docvqa}, infographics~\cite{masry2022chartqa,mathew2022infographicvqa} and UIs~\cite{chen2021websrc,hsiao2022screenqa,liu2024visualwebbench,wang2021screen2words,li2020widgetcaptioninggeneratingnatural}. However, all of these tasks consist of single-image or multi-image setups that contain the relevant information on at most one of the images. Current SoTA multimodal LLMs achieve impressive performance on these tasks, averaging >85\% zero-shot performance across UI and document related datasets like ChartQA~\cite{masry2022chartqa}, DocVQA~\cite{mathew2021docvqa}, InfographicVQA~\cite{mathew2022infographicvqa}, ScreenQA~\cite{hsiao2022screenqa} and WebSRC~\cite{chen2021websrc}. Motivated by the promise of multimodal LLMs to tackle various UI based single-page tasks, the research community has created a variety of multi page benchmarks focusing on UI Automation~\cite{deng2023mind2web,koh2024visualwebarena}, completing multi-step tasks on UIs based on user goals. A significant majority of these benchmarks require the use of a single website to complete the given task. In those containing multi-website tasks, typically each example contains one website per category~(e.g., shopping, restaurants, movies, education etc) but do not include multiple websites with similar functionality.

Real world web usage necessitates combining and comparing information from multiple webpages. When provided with a set of semantically related webpages, answering challenging questions requires a model to discern relevant information among confounding concepts presented with varied layouts and spatial structures. It can therefore be limiting to develop and evaluate web agents using only the standard QA formulation found in various UI automation datasets. Motivated to address this gap, we introduce WebQuest, a multi-page screen question-answering dataset, to assess information retrieval, reasoning and navigation capabilities of multimodal models. 

\begin{figure*}[h]
    \centering
    \includegraphics[width=\textwidth]{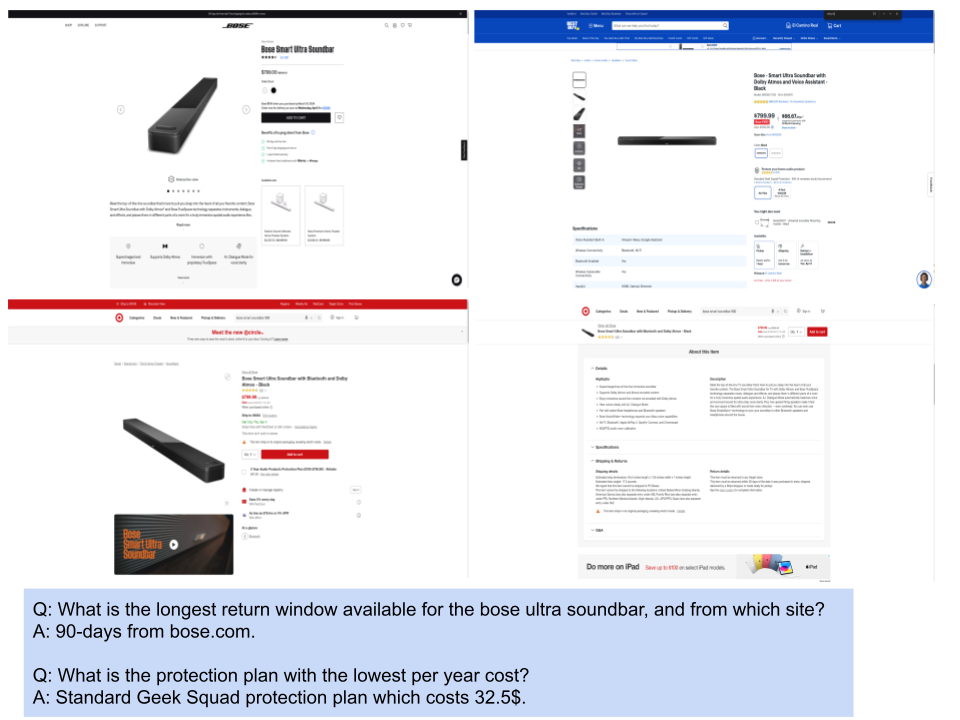}
    \caption{An example of multi-page question and answering in WebQuest. In this example, there are 4 screenshots and 2 question-answer pairs. Answering both questions needs reasoning over information extracted from the different screenshots.}
    \label{fig:example_of_multi_page_webquest}
\end{figure*}

We reformulate the QA task used in existing UI datasets in two ways. Firstly, we include questions requiring aggregation/reasoning across information extracted from multiple parts of the screen. Secondly, our dataset includes questions spanning multiple webpages and websites, which, we believe, more accurately reflects a user's ability and need to compare and reason across pages. In figure~\ref{fig:example_of_multi_page_webquest}, we provide an example from the dataset, illustrating the same product presented in different websites. In this case, the ability to reason over information extracted from multiple webpages is essential to compare the different offerings and make an informed choice. We acknowledge that some sophisticated websites or aggregator websites provide filters and comparison functionalities, but they are limited to popular options and do not always expose all the functionality available to the user. 

To allow for a detailed benchmarking of various models, WebQuest includes three question categories of increasing difficulty: single-screen QA, multi-screen QA, and QA on navigation traces. The first two involve multimodal content from one or more screens, while trace QA is based on sequences of screens navigated by humans to perform routine tasks. Figure~\ref{fig:example_of_multi_page_webquest} illustrates an example of multi-screen question answering.
\noindent Finally, we evaluate various state-of-the-art proprietary models GPT-4V~\cite{openai2024gpt4}, Gemini Flash~\cite{geminiteam2024gemini}, Claude 3~\cite{anthropic2024claude3}, and open source models like InstructBLIP~\cite{dai2023instructblip}, PaLIGemma~\cite{beyer2024paligemma} via prompt engineering and Chain-of-thought prompting, revealing a significant gap between single-page and multi-page reasoning.

 The main contributions of our work are multifold. We believe it will advance the field of UI understanding and automation.
\begin{itemize}
    \item
        WebQuest is the first dataset containing QA on UIs requiring reasoning across various parts of a single UI or multiple UIs.
    \item
        WebQuest is the first dataset with QA over long interaction sequences focused on information extraction and complex reasoning across the sequences.
    \item
        WebQuest includes 3 subsets of data containing single-screen, multi-screen and navigation traces, enabling detailed benchmarking of the capabilities of different models.
\end{itemize}

\section{Related Work}
\label{sec:related_work}
\noindent WebQuest is a dataset for Visual Question Answering across multiple images on web screens. In addition, it also includes navigation traces composed of the sequence of steps taken to arrive at the relevant screens to answer the questions. We present prior research along the different aspects covered by our dataset below.

\paragraph{Visual Question Answering}
The problem of answering questions based on the contents of an image has been studied extensively by the research community from VQA~\cite{antol2015vqa} on natural images, to TextVQA~\cite{singh2019towards} of text in natural images and DocVQA~\cite{mathew2021docvqa} containing documents with text, tabular structures and figures. Focusing more on documents and UIs, datasets like ChartQA~\cite{masry2022chartqa} and InfographicVQA~\cite{mathew2022infographicvqa} emphasize complex numerical and compositional reasoning.  
For web and UI understanding, QA datasets like WebSRC~\cite{chen2021websrc} focused on desktop screens and ScreenQA~\cite{hsiao2022screenqa} focused on mobile app screens; both contain questions which require extracting relevant information from a single screen. While related, our dataset contains questions that necessitate more complex reasoning and also span multiple screens. Similar to our dataset in having multiple images as input, Multi-page~DocVQA~\cite{tito2023hierarchical} requires visual extraction from up to 20 pages to answer questions. However, Multi-page~DocVQA requires only a single page that contains the answer to be identified whereas in WebQuest, the answer spans multiple screens. To the best of our understanding, there are no QA benchmarks spanning multiple websites or images in a task-oriented sequence, where a subset of pages are necessary for joint reasoning.
Other datasets like TaT-DQA~\cite{zhu2022towards} and DUDE~\cite{vanlandeghem2023documentunderstandingdatasetevaluation} contain questions over multiple pages, but are based on documents with tables and not rich UIs.

\paragraph{Visually-situated Grounding} Computer control and workflow automation involve decision-making grounded in visual inputs. Agents have to plan across diverse APIs, interfaces to execute the correct action at the OS and application level. Datasets either captures grounding in a multi-task setting via natural language expressions~\cite{liu2024visualwebbench} or end-to-end~\cite{rawles2024androidinthewild}. Previous modeling focused on text-based abstractions, while pixel-inputs are increasingly explored in recent work like \cite{shaw2023pixels}, \cite{yang2023set}, and \cite{zheng2024seeact}. Despite the increase in datasets related to single-page action grounding and multi-step navigation, our dataset is the first to focus on question answering in a single and multi page setting.

\paragraph{Web Navigation and Agent Benchmarks}
Recently datasets that focus on performing various tasks in web and mobile scenarios have
received widespread attention. Earlier efforts introduced simulated web and mobile environments, such as MiniWob++~\cite{liu2018reinforcement}, MoTIF~\cite{burns2022motifvln}, Mind2Web~\cite{deng2023mind2web} and
WebShop~\cite{yao2023webshop}. To facilitate autonomous web agents, WebArena~\cite{zhou2024webarena}, VisualWebArena~\cite{koh2024visualwebarena} and WebLinx~\cite{lv2024weblinx} pair interactive instructions with DOM trees, HTML and pixel-based environments. A majority of these datasets do not have step-by-step supervision for each action taken as part of completing the task. Recently released MMInA~\cite{zhang2024mmina} is the first to collect multi-hop, sequential navigation across websites. While they focus on long-range reasoning with only one website per task category (e.g. shopping, travel), WebQuest has multiple websites for each task which is closer to real world workflows. %

\section{WebQuest Benchmark}
\label{sec:data_collection_and_stats}

\begin{figure}[h!]
    \centering
    \begin{minipage}{0.45\textwidth}
        \centering
        \includegraphics[width=.9\textwidth]{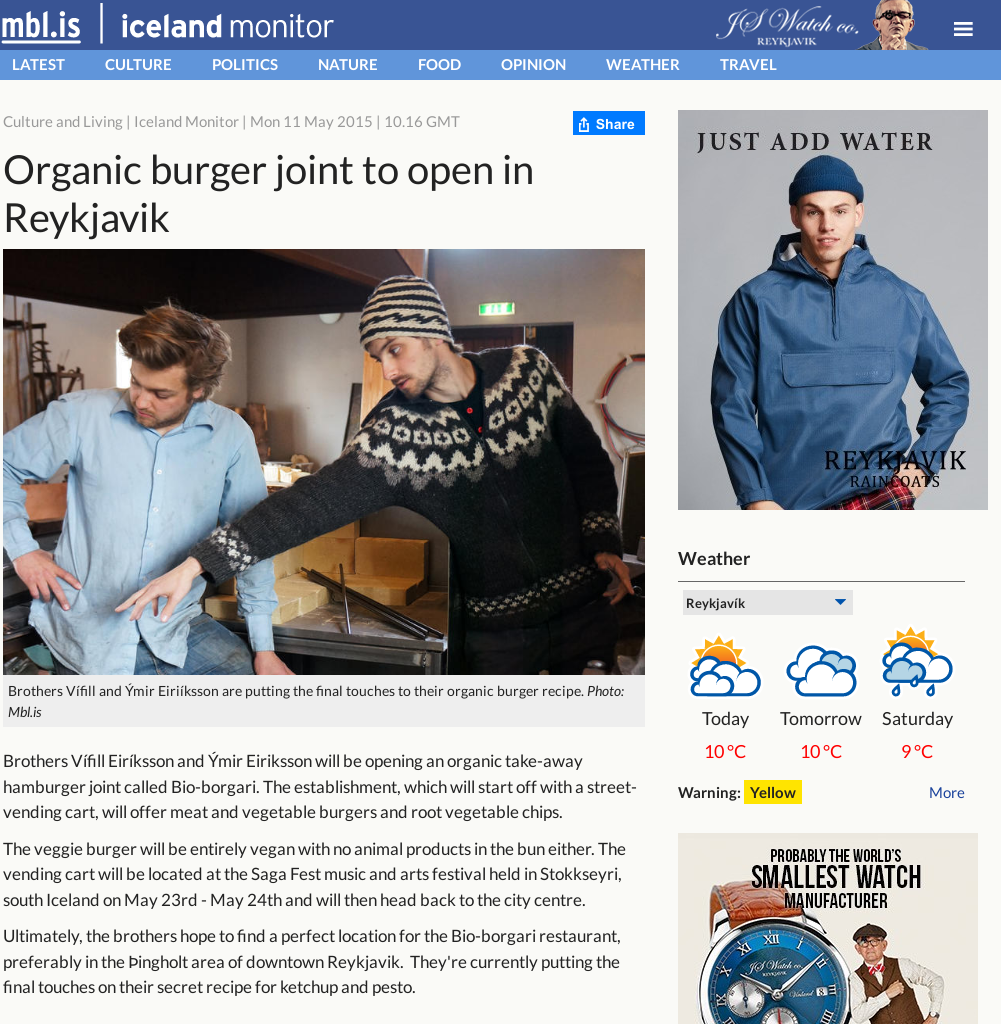}\\
                \begin{minipage}[b]{\textwidth}
                \small
                \texttt{
                    \\
                    \textbf{Question:} How many more days will it be cloudy than rainy?	\\
                    \textbf{Answer:} 1
                }
            \end{minipage}
        \caption{An example of Single Screen QA, where the task is to count how many cloudy days before a rainy one, and the weather conditions are depicted by pictograms. }
        \label{fig:single_page_iceland}
    \end{minipage}
    \hspace{1.5mm}
    \begin{minipage}{0.45\textwidth}
        \centering
        \includegraphics[width=\textwidth]{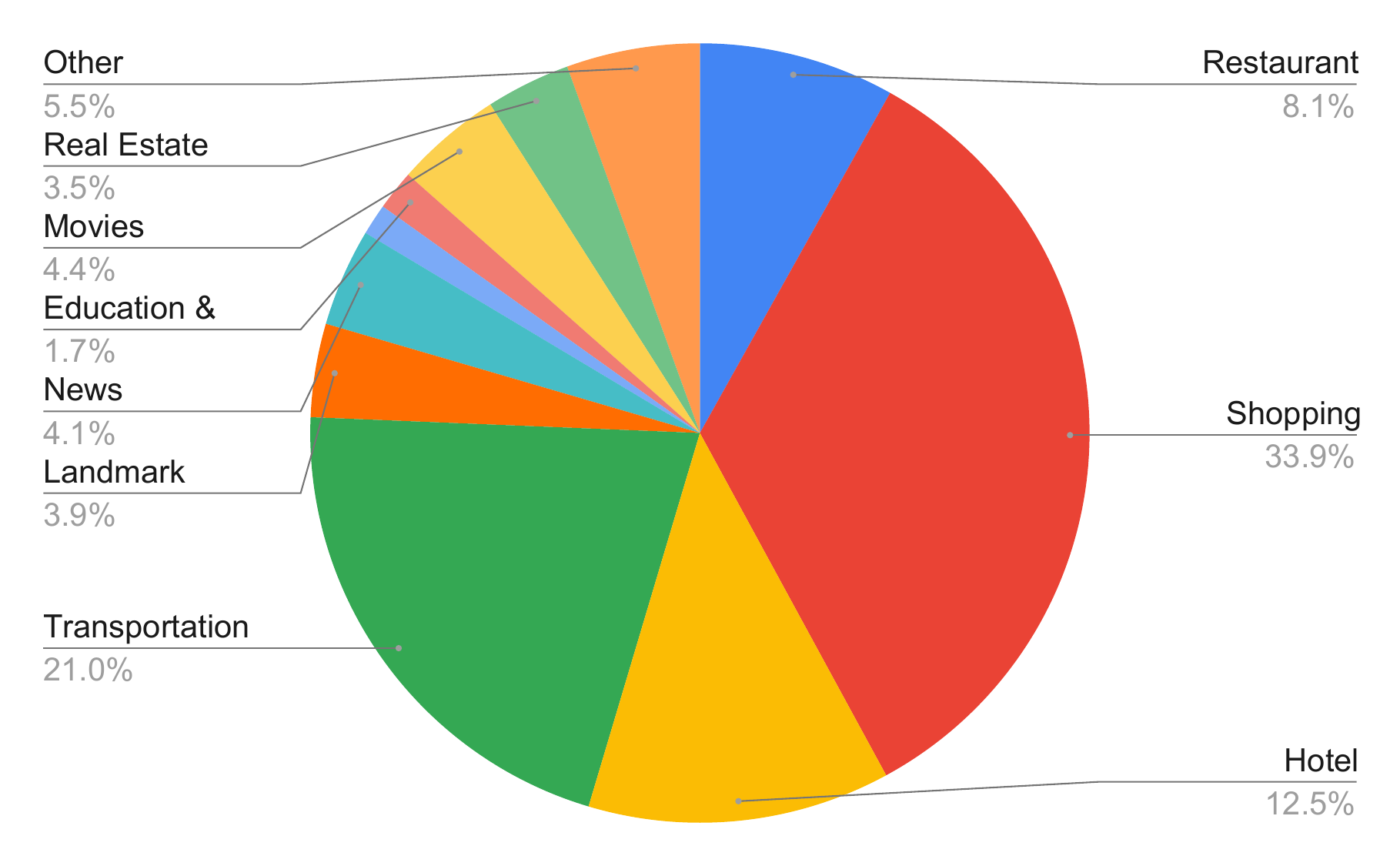}
        \caption{Distribution of website categories of Single Screen QA examples.}
        \label{fig:single_screen_category}
    \end{minipage}
\end{figure}

\begin{figure}[h!]
    \centering
    \includegraphics[width=.6\textwidth]{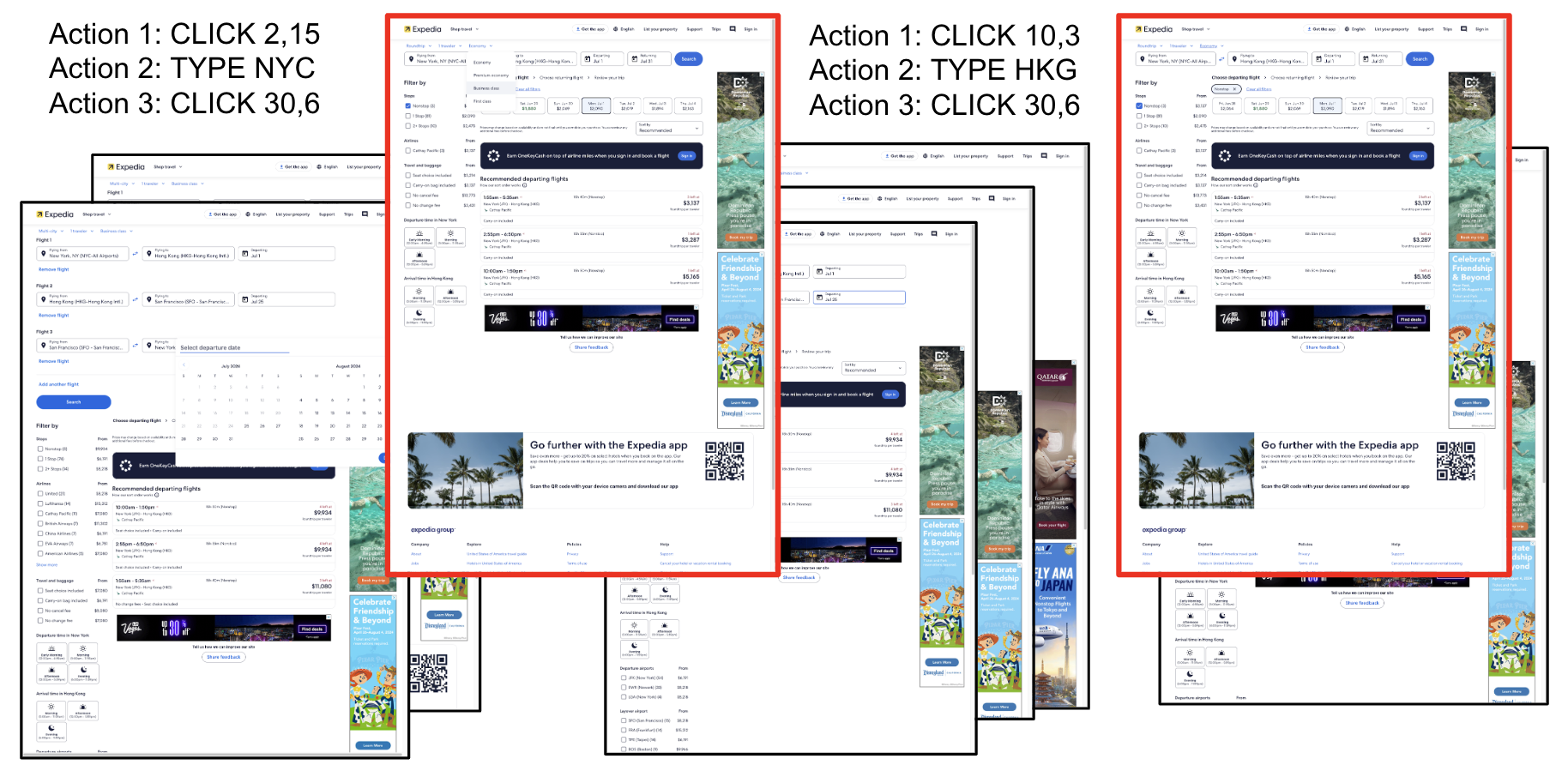}\\
    \begin{minipage}{.6\textwidth}
        \small
        \texttt{
            \\
            \textbf{Question:} For nonstop flight from New York City to Hong Kong, how much more expensive is business class than economy?\\
            \textbf{Answer:} \$3054
        }
    \end{minipage}
    \caption{An example of Trace QA on airfare differences across cabin classes. A browsing session is provided with all screen sequences, but not all of them are required to answer the question. }
    \label{fig:trace_qa_airfare_difference}
\end{figure}
\begin{figure}[h!]
    \centering
    \includegraphics[width=.95\textwidth]{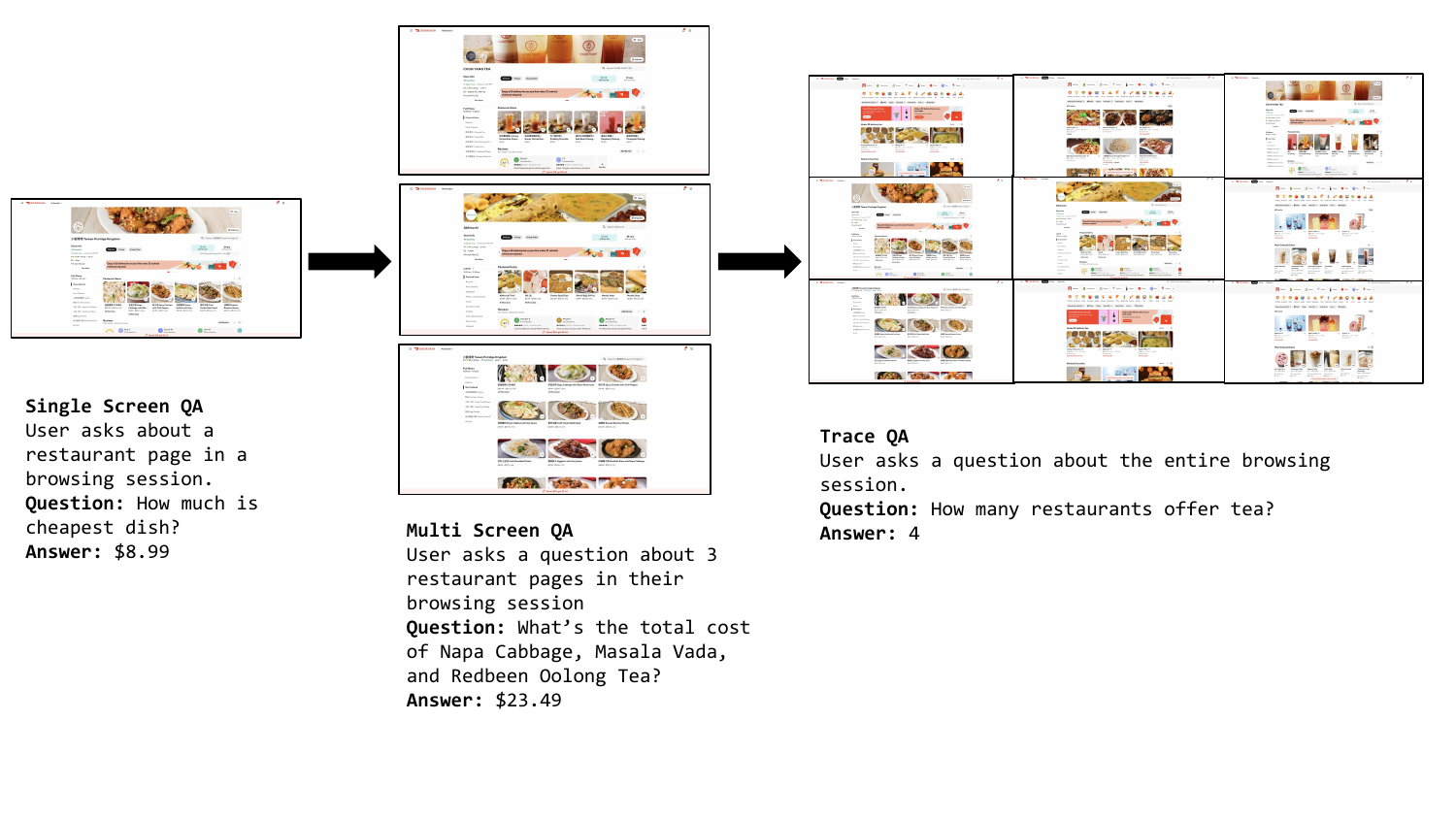}\\
    \caption{This figure demonstrates the relationships among Single Screen QA, Multi Screen QA, and Trace QA. Single Screen QA focuses on a single page, Multi Screen QA focuses on multiple pages within a browsing session, and Trace QA focuses on the entire browsing session. }
    \label{fig:compare_each_qa_dataset}
\end{figure}

WebQuest is a multimodal benchmark focused on question answering based on the contents of web UIs. It is targeted to capture the scenarios where users navigate to different websites to retrieve information. In contrast with many question answering datasets which are based on information contained in a single image/UI, WebQuest enhances the scope of the QA task by requiring the combination of information from multiple distinct parts of UIs. Furthermore, WebQuest also contains questions spanning multiple related webpages and websites. To better understand the capabilities and characterize the performance of various modeling approaches, we split the dataset into three categories of increasing difficulty:
\begin{enumerate}
    \item \emph{Single screen QA}: Questions based on the contents of a single screen~(see Figure~\ref{fig:single_page_iceland} for an example);
    \item \emph{Multi screen QA}: Questions spanning multiple semantically related web pages. For this task, the web pages are filtered such that all of them are required to answer the question of interest~(see Figure~\ref{fig:example_of_multi_page_webquest} for an  example);
    \item \emph{Trace QA}: Questions about the entire sequence of screens viewed by the user as they navigate the web with a particular goal~(see Figure~\ref{fig:trace_qa_airfare_difference} for an example). 
\end{enumerate}
Please refer to figure~\ref{fig:compare_each_qa_dataset} for an overview of the relation between the different categories. In the remainder of this section, we describe the motivation, details of the different categories in WebQuest and the data collection process. To arrive at the source websites, we initially curate a list of categories based on their functionality and then compile a list of popular websites in each category.

\subsection{Single screen QA}
\label{sec:single-screen-qa}
This category contains question answer pairs where each question is based on the contents of a single web UI. Compared to other UI based QA datasets~\cite{chen2021websrc,hsiao2022screenqa,liu2024visualwebbench}, answering each question requires extracting information from different parts of the UI and then combining them using mathematical or logical operations. Some of the common cases covered in this splits include counting the number of entities which meet a criteria~(e.g., number of items below a certain price), aggregating information etc.
This category consists of 542 examples. To collect this dataset, raters start with the list of popular websites provided to them and are asked to follow the instructions:
\begin{enumerate}
    \item Explore the website and think about what the website is used for and how users typically interact with it.
    \item Consider the kind of questions users might be interested in when using the website. Focus on questions using information from multiple parts of the screen and needing arithmetic or logical reasoning.
    \item Take a screenshot and generate a relevant QA pair.
\end{enumerate}

During the data collection process, the authors of this work frequently reviewed the generated examples to ensure that the question-answer pairs are of high quality, diverse and reflect real world scenarios. Additionally, two rounds of data validation were performed to remove screenshots that contain PII information. Figure~\ref{fig:single_screen_category} shows the distribution of different website categories in the Single screen QA dataset. We can observe that the Single screen QA dataset contains websites across many categories with shopping and restaurants being the most common ones. The detailed statistics are listed in Appendix section B.

\begin{figure}[h!]
    \centering
    \begin{minipage}{0.45\textwidth}
        \centering
        \includegraphics[width=\textwidth]{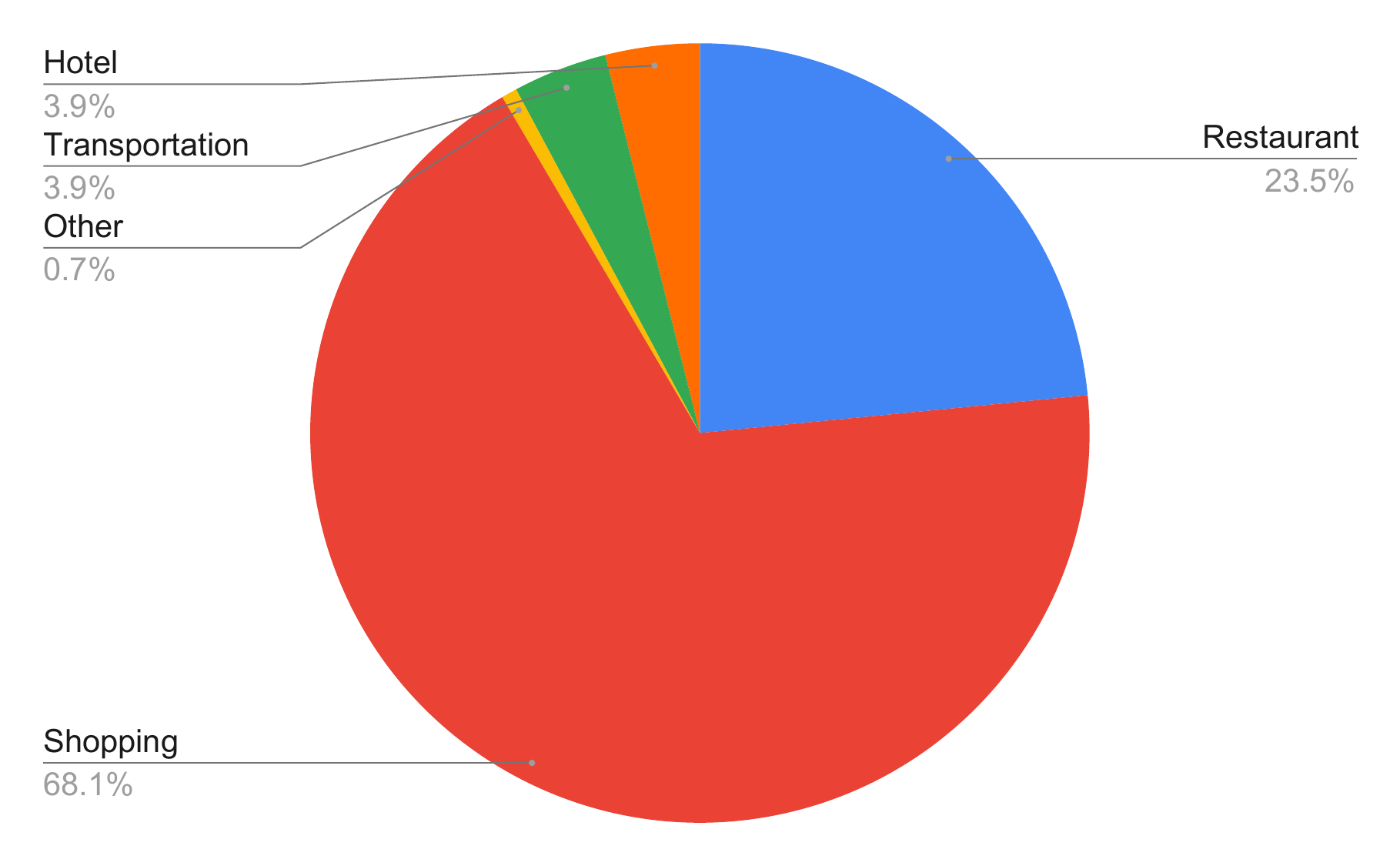}
        \caption{Distribution of categories of Multi Screen QA examples. The websites span over 5 categories with a higher proportion in Shopping and Restaurants.}
        \label{fig:multi_screen_category}
    \end{minipage}
    \hspace{1mm}
    \begin{minipage}{0.45\textwidth}
        \centering
        \includegraphics[width=\textwidth]{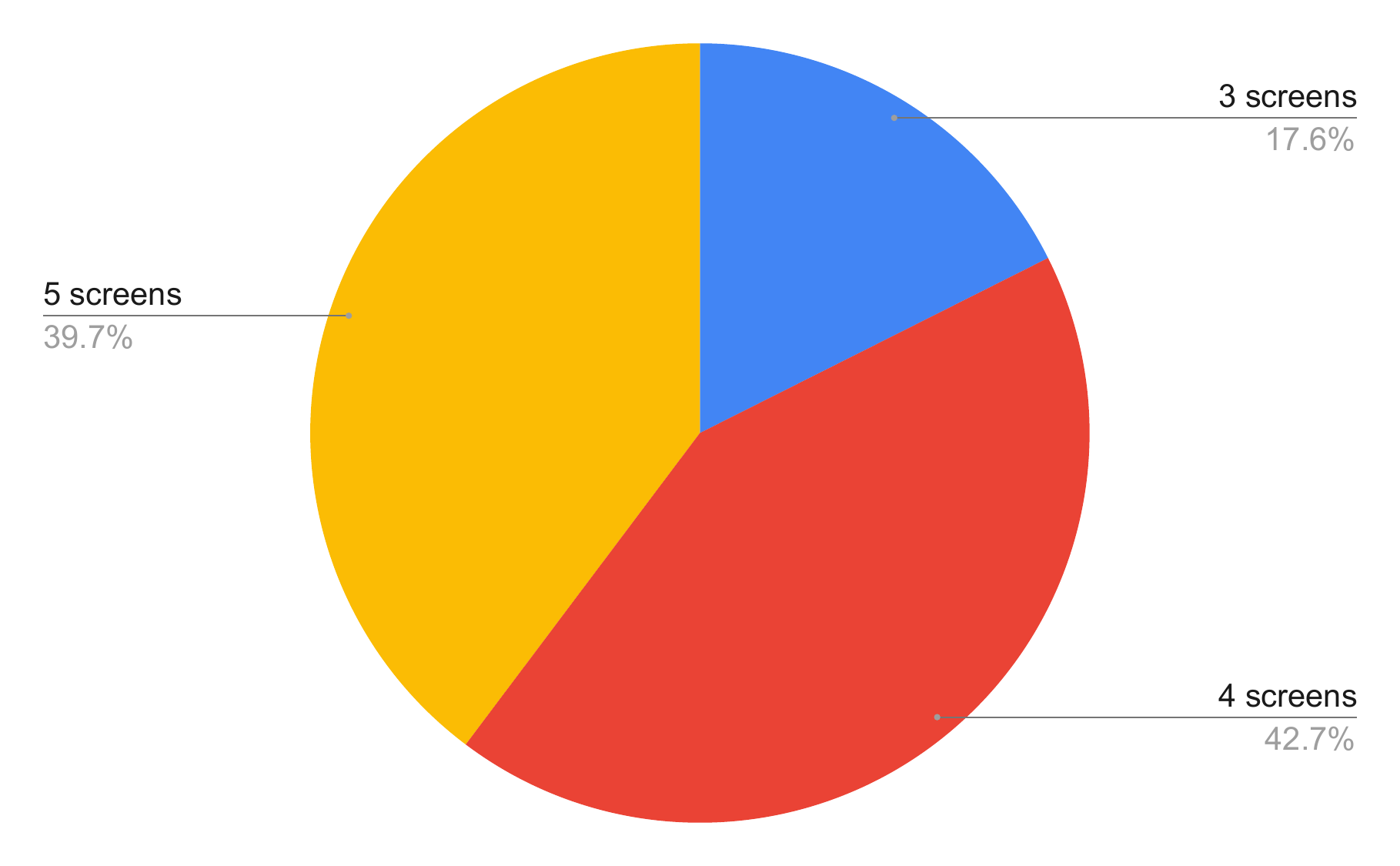}
        \caption{Distribution of number of screens for Multi Screen QA examples.}
        \label{fig:multi_screen_num_image}
    \end{minipage}
\end{figure}

\subsection{Multi screen QA}
\label{sec:multi-screen-qa}
In the Multi Screen QA task, we aim to capture instances where users are browsing the web with the aim of gathering information focused on a particular task, e.g., shopping for a sound bar as depicted in fig~\ref{sec:multi-screen-qa}. This could involve tasks like comparing product prices/attributes across websites, evaluating options for travel, filtering feasible options for movies etc. The overall goal is similar to the Trace QA described in sec~\ref{sec:trace-qa}, however for this case, the screens containing relevant information for answering the questions are already filtered. This occurs in cases where a user has explored several relevant webpages, opened them in different tabs or taken screenshots, and then seeks answers pertaining to these pages. It is important to note that the different screens in each QA pair are semantically related and can come from webpages in the same website or alternative websites offering similar functionality.

This category contains 307 QA pairs each encompassing multiple web screens. The overall guidance provided to raters is similar to the one in section~\ref{sec:single-screen-qa}. One major difference is that now the raters are instructed to capture multiple screenshots from different webpages. To ensure that related websites are explored for each QA pair, raters are provided a curated list of popular websites grouped into categories. This curated list serves as a guide for selecting coherent screens for each QA pair. Each rater is first assigned a category for each question, and are asked to follow the instructions:
\begin{enumerate}
    \item Choose 2-5 websites from the assigned category.
    \item Imagine a user researching a product, service, or location online, comparing information across webpages and websites.
    \item Take 3-5 screenshots from websites related to the chosen use case.
    \item Formulate a question a user might have based on the combined information in the screenshots.
    \item Ensure that the question requires logical or arithmetic operations and needs information from all screenshots to answer.
    \item Provide a short answer to the question.
\end{enumerate}
The exact instructions given to the raters are specified in the appendix. \\

The authors of this work periodically review the data quality and provide feedback to the raters. Furthermore, 2 rounds of data validation were performed to prevent inappropriate and unsafe screens. Figures~\ref{fig:multi_screen_category}, ~\ref{fig:multi_screen_num_image} contain the distribution of different website categories and number of screens per example. Detailed statistics for the categories are presented in section B of the Appendix.

\subsection{Trace QA}
\label{sec:trace-qa}
In the Trace QA category, similar to the Multi Screen QA category described in section~\ref{sec:multi-screen-qa}, we aim to capture instances where users are browsing the web with the aim of gathering information focused on a particular task. In the rest of this section, we use the term browsing session to represent the different webpages visited as part of completing this task. All of the webpages encountered during the browsing session are associated with the corresponding QA pair. 
However, do note that only a subset of the screens are relevant to answering the question. 
The main difference between this category and the \emph{Multi screen QA} category is that, in \emph{Multi Screen QA} each QA pair is associated with only the screens containing information relevant to answer the question, whereas in \emph{Trace QA} each QA pair contains the different webpages in the browsing setting. We consider \emph{Multi screen QA} as a simplified version of the \emph{Trace QA} task which enables us to measure the capabilities of different modeling approaches on answering questions across multiple screens.

\noindent For the data collection, similar to the multi-screen screen QA data collection, we provide top websites and their associated categories to the human raters. The raters browse through websites and record screen traces. Below are the instructions provided to the raters:
For the \emph{Trace QA} split, similar to the \emph{Multi Screen QA} split, the raters are asked to focus on specific website category. 
The raters install custom Chrome Extension which enables them to record their interactions on the screen. The chrome extension is meant to be turned on when the rater is ready to collect data. They can pause, review or delete any of the data captured. Each rater is first assigned a category for each question, and are asked to follow the instructions:
\begin{enumerate}
\item Imagine a user researching a product, service, or location online, comparing information across websites and pages. Consider the kind of questions the user might be interested to know.
\item Go to the first web page and start the trace recording.
\item Navigate through the different webpages as part of the trace. Once the navigation is complete, click "stop" on the recording and review the screenshots captured.
\item Generate the relevant question.
\item Ensure that the question focuses on logical reasoning and arithmetic.
\item Ensure that answering the question needs to use information from about 3-5 screenshots.
\end{enumerate}

\begin{figure}[h!]
    \centering
    \begin{minipage}{0.45\textwidth}
        \centering
        \includegraphics[width=\textwidth]{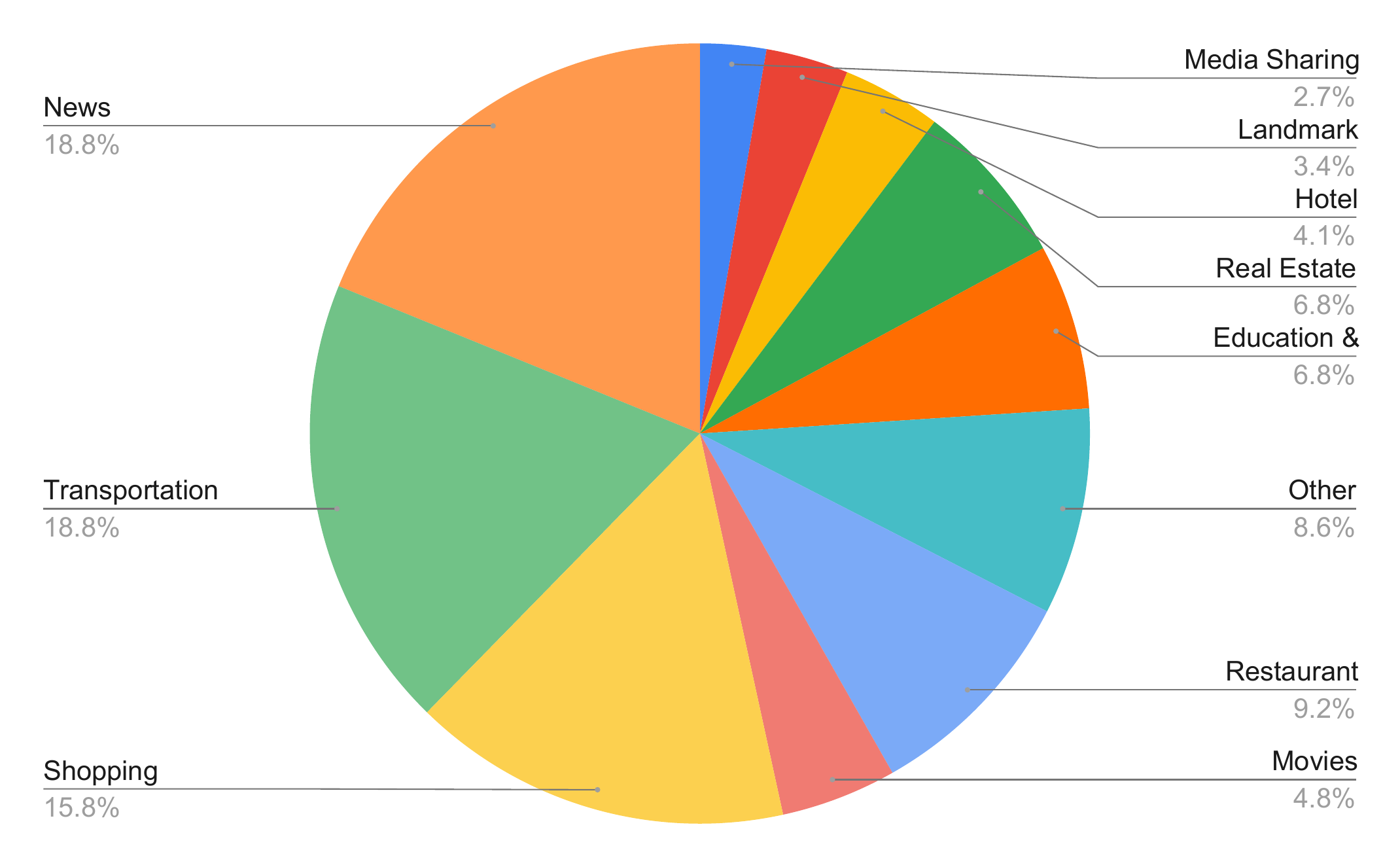}
        \caption{Distribution of categories of Trace QA examples. News, transportation and shopping categories contain the most number of traces.}
        \label{fig:trace_category_pie}
    \end{minipage}
    \hspace{2mm}
    \begin{minipage}{0.45\textwidth}
        \centering
        \includegraphics[width=\textwidth]{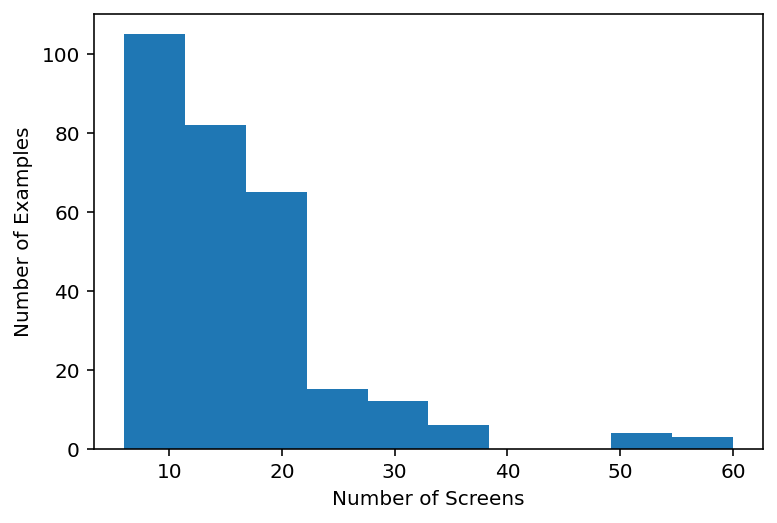}
        \caption{Distribution of the length of traces for the Trace QA benchmark. There are a few very long traces with more than 50 screens, while a majority of the traces contain fewer than 20 screens.}
        \label{fig:trace_num_image}
    \end{minipage}
\end{figure}

We display the distribution of the number of screens in Figure~\ref{fig:trace_num_image},
and the distribution in different categories in Figure~\ref{fig:trace_category_pie}. On average, each trace contains 16 screens. Please see Table~\ref{tab:stats_trace} in the Appendix for exact counts of the websites for the different questions. The data is validated to ensure no PII is present and the data is manually reviewed for correctness.

\subsection{Rater Details}
For the \emph{Single Screen QA} and \emph{Multi Screen QA} tasks, we used 10 raters with ages between 25 and 35 based in India. The raters were employed as contractors to help with a variety of data collection tasks apart from this task.

For the \emph{Trace QA} task, the authors of the paper collected the data. They are based in the United States and Europe.

\section{Experiments}
\label{sec:experiments}
\subsection{Metrics}
To evaluate the performance of various models on the WebQuest benchmark, we utilize a variant of the Relaxed Accuracy metric used for the ChartQA~\cite{masry2022chartqa} dataset. It is noteworthy that over half of the answers in this benchmark are numerical values. For these numerical value answers, we use heuristics to extract the numbers from predictions and ground truths. For floating-point answers, similar to Relaxed Accuracy, we allow a margin of ±0.05 in predictions to be considered correct. In contrast, for integer and string answers, we use SQuaD~\cite{rajpurkar2016squad} pre-processing and exact match between the prediction and ground truth as the success criterion. Furthermore, we incorporate multiple ground truth variants generated using Gemini Pro~\cite{geminiteam2024gemini} to accommodate various acceptable answer formats and consider the final score as the maximum score across all ground truth variants.
\subsection{Baselines}
We evaluate the performance of a number of state-of-the-art MLLMs and leading open source models on our benchmark. 
We use pretrained parameters and default settings of API-based models for evaluation. The model input includes the screen image(s) and question. Although image ordering should not be relevant to the questions, we use the same ordering of images for Multi Screen QA and the original navigation sequence orders for Trace QA.
For models we used out of the box, inference time techniques like prompt engineering and Chain-of-thought prompting~\cite{wei2022chain} were utilized to improve their performance. In our Chain-of-thought~\cite{wei2022chain} prompt, we include stages (steps) to analyze the screenshot and question, extract information from each screen, analyze extracted information, then finally to generate the answer in short phrases. We evaluated various state of the art models like
GPT-4V~\cite{openai2024gpt4},
Gemini Flash~\cite{geminiteam2024gemini},
Claude 3~\cite{anthropic2024claude3}. For Claude 3~\cite{anthropic2024claude3} models, the model input is limited to 20 images. Therefore for Trace QA evaluations, we only use the last 20 images in the sequence. Furthermore, due to memory constraints for Claude 3~\cite{anthropic2024claude3} models, we resize the images preserving aspect ratio such that the shorter side is 1440 pixels for Trace QA, whereas we use the original image resolutions for Single Screen QA and Multi Screen QA. For GPT-4V~\cite{openai2024gpt4} and Gemini Flash~\cite{geminiteam2024gemini} evals we use the original image resolutions for all questions.
We also evaluate open sourced models, including BLIP2~\cite{li2023blip2bootstrappinglanguageimagepretraining}, InstructBLIP~\cite{dai2023instructblip} and PaLIGemma~\cite{beyer2024paligemma}. We evaluate BLIP2 and InstructBLIP in a zero-shot setting, and in both cases, use the variant based on Flan-T5-XXL language model~\cite{chung2022scalinginstructionfinetunedlanguagemodels}. Following the recommendations from the official documentation~\cite{googleai2024paligemmadocs}, we evaluate a fine-tuned version of PaliGemma. We fine-tune the pre-trained \emph{pt-896} checkpoint on a mixture of open source datasets: DocVQA~\cite{mathew2021docvqa}, mobile ScreenQA~\cite{hsiao2022screenqa}, ChartQA~\cite{masry2022chartqa} and InfographicVQA~\cite{mathew2022infographicvqa}. We set 1e-5 as the initial learning rate, with a warmup of 0.05\% and a cosine decay. We run fine-tuning for 10 epochs.

\begin{table*}
\centering
\caption{Performance of various MLLMs on variants of the \emph{WebQuest benchmark}. The metric used here is Relaxed Accuracy. All these models, except \emph{PaLIGemma} are used in a zero shot setting without any fine-tuning. For each of the models, prompt tuning was performed separately. Overall, we can see that all of the MLLMs perform a lot better on the \emph{Single Screen QA} split and their performance decreases as the number of screens increases. Also, note that the performance of the MLLMs on the \emph{Trace QA} split is considerably lower than on the \emph{Multi Screen QA} split.}
\label{tab:all_results}
\begin{tabular}{l r r r r r r}
\toprule
\textbf{Model} & \multicolumn{2}{c}{\textbf{SingleScreen QA}} & \multicolumn{2}{c}{\textbf{MultiScreen QA}} & \multicolumn{2}{c}{\textbf{Trace QA}}  \\
& no CoT & CoT & no CoT & CoT & no CoT & CoT  \\
\midrule
\multicolumn{7}{c}{Proprietary models} \\
\midrule
Claude-3 Sonnet & 27.3 & 38.9 & 15.4 & 31.2 & 23.2 & 19.6 \\
Claude-3 Opus & 25.3 & 40.4 & 16.3 & 23.4 & 18.5 & 26.6 \\
Gemini Flash & 34.9 & 48.9 & 15.8 & 31.1 & 24.6 & 35.9 \\
GPT-4V & \textbf{36.7} & \textbf{57.4} & \textbf{23.2} & \textbf{44.0} & \textbf{29.0} & \textbf{40.8} \\
\midrule
\multicolumn{7}{c}{Open-source models} \\
\midrule
PaLIGemma & 10.5 & N/A & <5\% & N/A & N/A & N/A \\
BLIP-2 & 6.7 & N/A & N/A & N/A & N/A & N/A \\
InstructBLIP & 9.1 & N/A & 8.8 & N/A & N/A & N/A \\
\bottomrule
\end{tabular}
\label{tab:all_results}
\end{table*}

\begin{figure}[h!]
    \centering
    \begin{minipage}{0.45\textwidth}
        \centering
        \includegraphics[width=\textwidth]{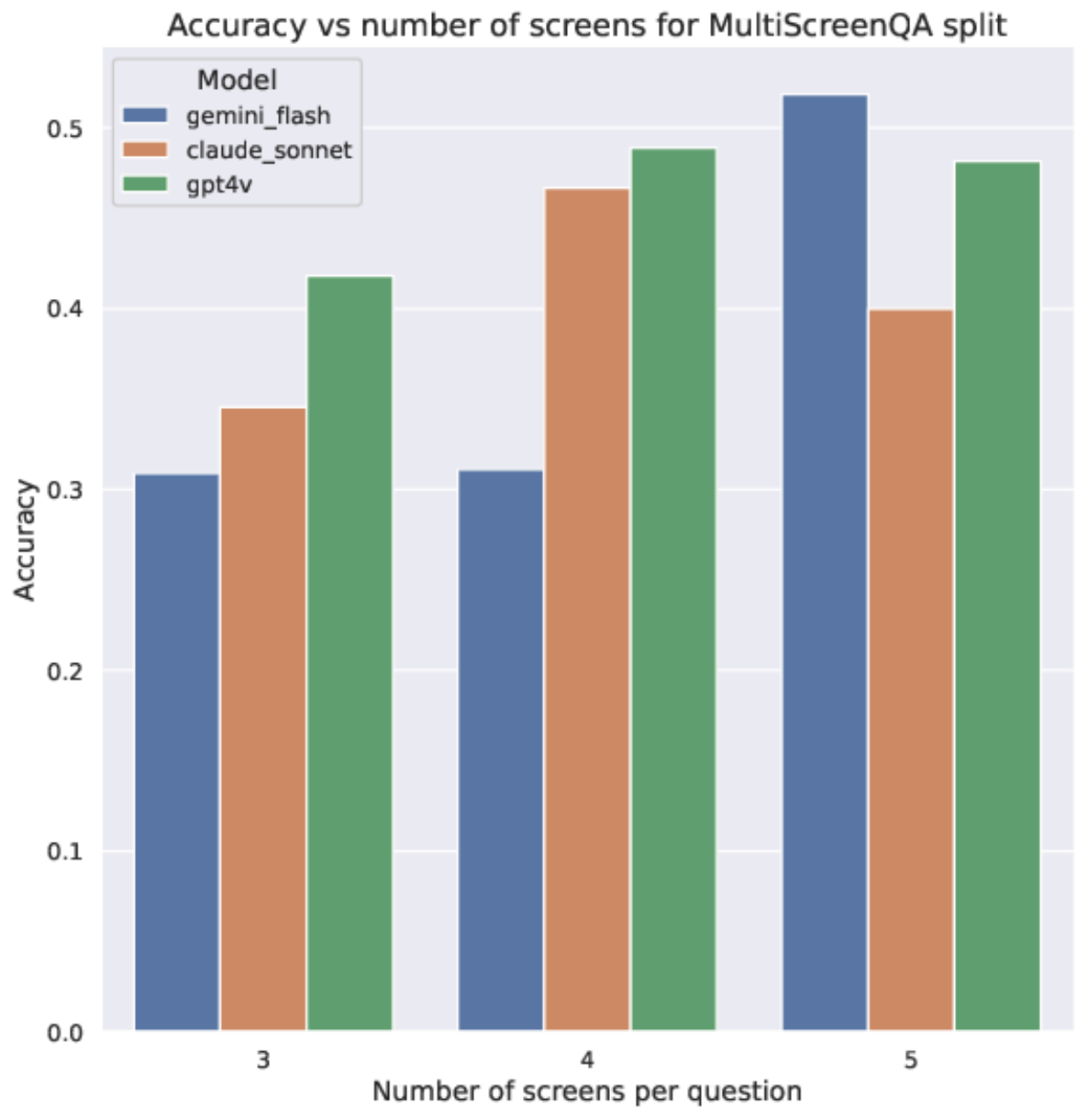}
        \caption{Performance of different models vs number of screens in each question. Perhaps surprisingly, we do not see clear evidence of model performance reducing as number of screens increases.}
        \label{fig:results-num-screens}
    \end{minipage}
    \hspace{2mm}
    \begin{minipage}{0.45\textwidth}
        \centering
        \includegraphics[width=\textwidth]{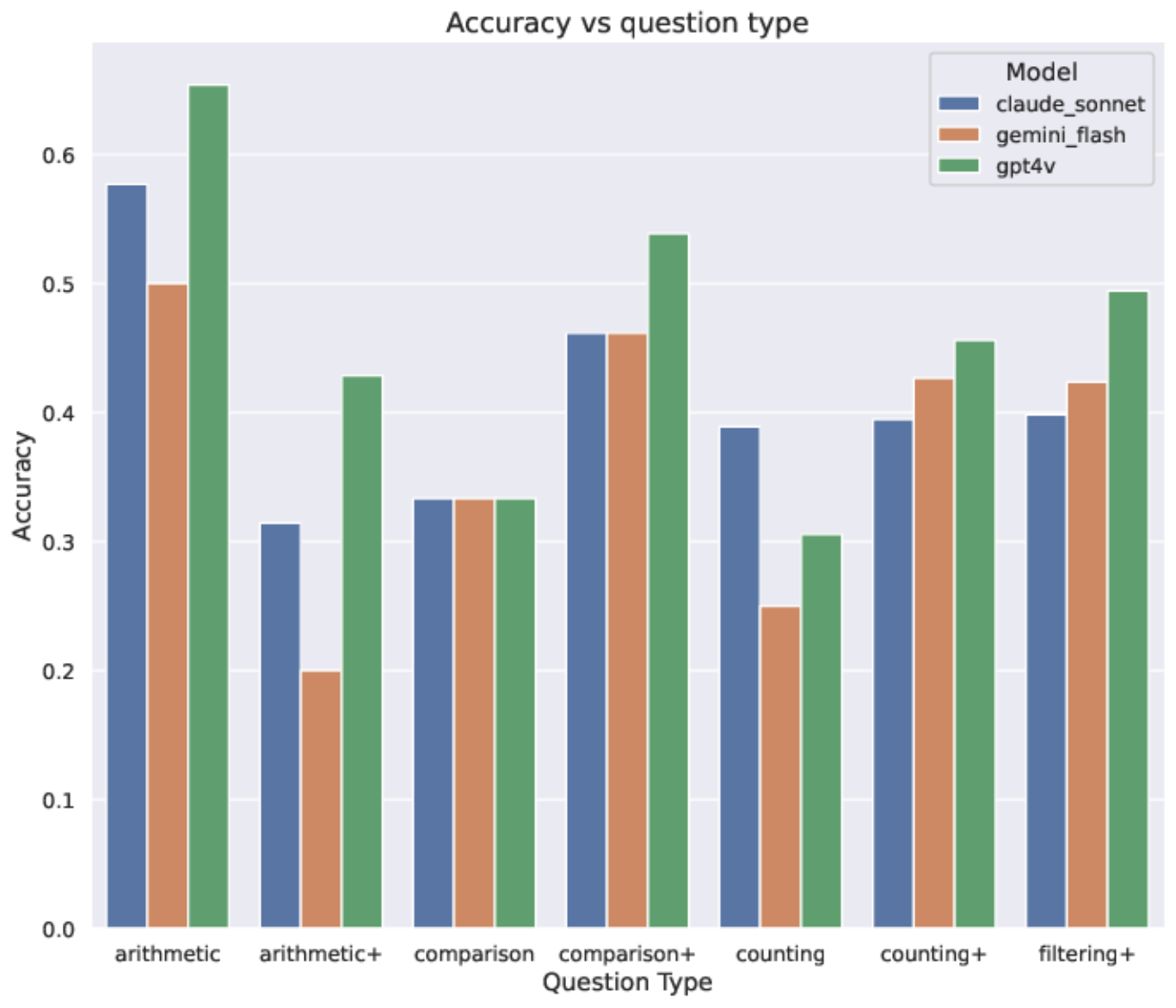}
        \caption{Performance of different models for different question types. The types ending in a \emph{+} sign e.g., \emph{arithmetic+} indicate questions which contain multiple types with arithmetic being one of them. Here we notice that the different models do not perform well on questions involving counting, whereas they do well on arithmetic questions. All models presented here were evaluated with CoT prompts.}
        \label{fig:results-question-type}
    \end{minipage}
\end{figure}

\subsection{Main Results}
In this section, we present a comprehensive comparison of different MLLMs on our WebQuest benchmark. To better understand the capabilities and failure cases of different models we present results on the various subsets described in section~\ref{sec:data_collection_and_stats}. The best performance for each MLLM on the WebQuest splits can be seen in table~\ref{tab:all_results}. We analyze the results in the sections below.

\subsubsection{Single Screen QA}
As described in section~\ref{sec:data_collection_and_stats}, the single screen question answering split involves answering a question based on the content of a single web screen. On this task, we notice that all the MLLMs perform relatively well compared to Multi Screen QA and Trace QA. Chain-of-thought~\cite{wei2022chain} prompting improves the scores on average 15.4\%. We also notice a significant performance difference between proprietary models and smaller open source models PaLIGemma and InstructBLIP.

\subsubsection{Multi Screen QA}
\label{sec:experiments_results_msqa}
To recap briefly, the multi screen question answering split involves question answering across various relevant screens from different websites. As we saw in section~\ref{sec:data_collection_and_stats}, the images are from related websites and on average, each QA pair contains 4.22 screens with 3, 4 or 5 screens. For this split, we notice that all the MLLMs perform worse than the single screen split. With some MLLMs, we notice considerable performance improvement with Chain-of-thought~\cite{wei2022chain} prompting and few shot evaluations. However, the performance for all the MLLMs drops to on average 32.4\% on this split, highlighting the difficulty some of these models have in reasoning across multiple images. We further analyze the performance gap in section~\ref{sec:experiments_analysis}. For open-source models, similar to Single Screen QA, we notice a significant performance gap to proprietary models. Also, we could not adapt BLIP2 model to run inference with multiple images, so we do not report any numbers for this split. As for PaLIGemma, since it was fine-tuned on single page VQA datasets (including DocVQA~\cite{mathew2021docvqa}, mobile ScreenQA~\cite{hsiao2022screenqa}, ChartQA~\cite{masry2022chartqa} and InfographicVQA~\cite{mathew2022infographicvqa}), the performance is very low. 

\subsubsection{Trace QA}
The main difference between this split and the Multi Screen QA~\ref{sec:experiments_results_msqa} is that this split also contains the screens seen during navigation to the relevant screens for answering the questions. As stated in section~\ref{sec:data_collection_and_stats}, the Multi Screen QA split is a simplified version of this split as it already contains only the relevant screens. This split can also be used for evaluating different agent systems since it contains action information for each step enabling step-by-step supervision compared to existing datasets. On average, each question has $15.8$
screens, out of which relevant information is present on $3.2$ screens.
We notice that the average performance of all the models is worse on this split compared to both the Single Screen and Multi Screen QA splits. With Chain-of-thought prompting~\cite{wei2022chain} and prompt engineering, MLLMs on average achieve a score of 30.6\%. Out of the MLLMs we evaluated, GPT-4V~\cite{openai2024gpt4} achieves the best performance with a score of 40.8\% on this dataset. Claude 3~\cite{anthropic2024claude3} models performs significantly worse than GPT-4V~\cite{openai2024gpt4} and Gemini Flash~\cite{geminiteam2024gemini} models, likely due to the extra image resizing step to avoid out of memory errors. For open-source models, the large number of input images per example resulted in out of memory errors.

\subsection{Analysis}
\label{sec:experiments_analysis}
In this section, we compare the performance of different MLLMs on WebQuest in Table~\ref{tab:all_results}. We highlight some of our significant findings below.
\subsubsection*{Single screen vs Multi screen}
We observe that all the models including the largest ones achieve much higher performance on single screen tasks compared to multi screen tasks. However, as we can see in figure~\ref{fig:results-num-screens}, for the \emph{MultiScreen QA} split, there is no clear evidence of model performance decreasing as the number of relevant screens increase. From figure~\ref{fig:results-question-type} we observe a significant performance gap in questions involving counting across screens compared to arithmetic or comparison operations.

\subsubsection{The impact of Chain-of-thought prompting}
We observe that Chain-of-thought~\cite{wei2022chain} prompting results in significant improvements across all models, even with the largest models across the GPT-4~\cite{openai2024gpt4}, Gemini~\cite{geminiteam2024gemini} and Claude 3~\cite{anthropic2024claude3} model families. On average, we see improvements of 15\% on \emph{Multi screen QA} split and 6.7\% on the \emph{Trace QA} split. When we manually inspect model step-by-step responses, we observe that the model is able to answer sub-questions on various screens correctly and this ability results in better overall performance. We also observed Chain-of-thought prompting~\cite{wei2022chain} improves multi-page to a larger extent than single-screen tasks (avg. 20.3\% vs 15.4\%) for most models. 
Claude-3 Opus~\cite{anthropic2024claude3}, which had the second highest baseline performance without Chain-of-thought prompting, ranked last among proprietary models when evaluated with CoT. One of the causes of reduced performance was that, with Chain-of-thought prompting, Claude-3 Opus produced N/A 11\% of the time in Multi-page QA. 

\subsubsection{Model performance on Chain-of-thought reasoning steps}
\begin{figure}[h!]
    \centering
    \includegraphics[width=.7\textwidth]{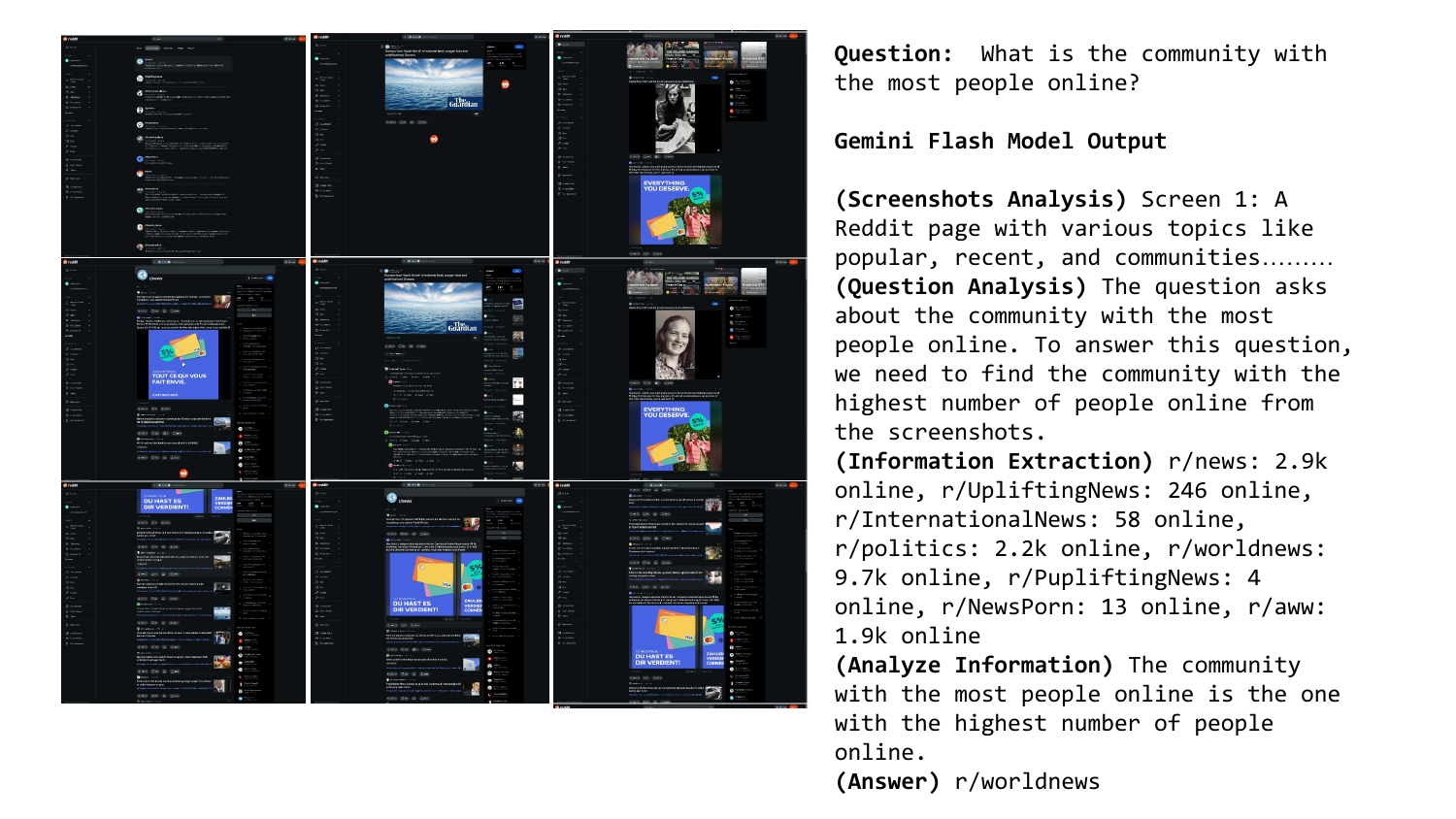}\\
    \caption{With Chain-of-thought prompting, Gemini Flash extracts relevant information from the screen sequence, then generates the correct answer. }
    \label{fig:trace_success_example}
\end{figure}

In our Chain-of-thought~\cite{wei2022chain} prompt, we split the original questions to 4 reasoning steps, including question analysis, screenshot analysis, information extraction, analyze information, and answer generation. 
We observe that all proprietary models perform well for question analysis, as illustrated in ~\ref{fig:trace_success_example}. For screenshot analysis, some models focus on higher-level semantics with common sense context while successful examples showed more attention to detail. i.e. "The second screenshot is a website for a restaurant called Vic's, showing popular menu items and their prices." vs. "Vic's restaurant with items like Cacio e Pepe for \$20.00, Roasted Chicken for \$28.00, and Rainbow Cookie Plate for \$8.00". Despite both models being successful in information extraction, the model that is more abstract in screenshot analysis performed the incorrect arithmetic and produced the wrong answer, and the correct model's screenshot analysis were more direct. Interestingly, we found some models to consistently do better at one and worse at the other step, but we observed no strong correlation between final answer correctness with any specific steps.

\begin{figure}[h!]
    \centering
    \begin{minipage}{0.48\textwidth}
        \centering
        \includegraphics[width=\textwidth]{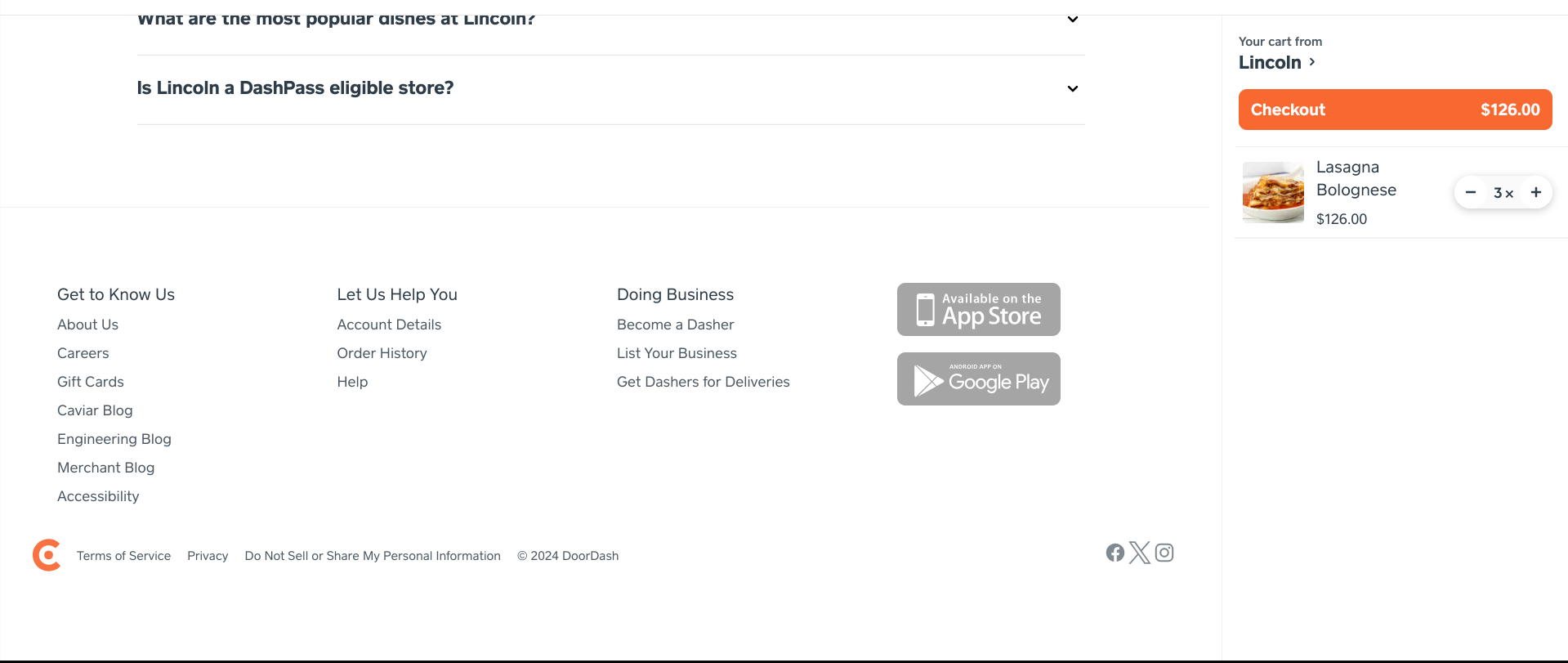}\\
        \includegraphics[width=\textwidth]{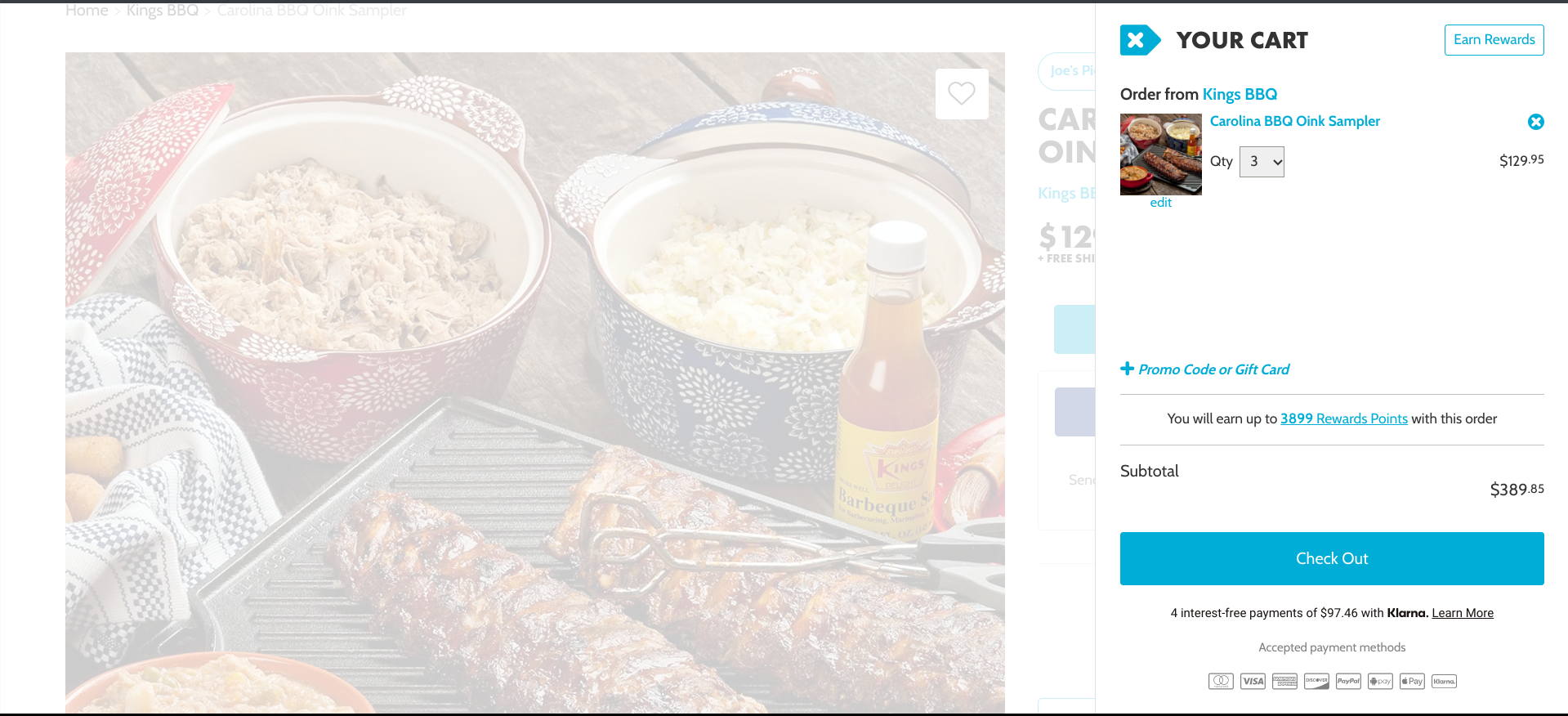}\\
        \includegraphics[width=\textwidth]{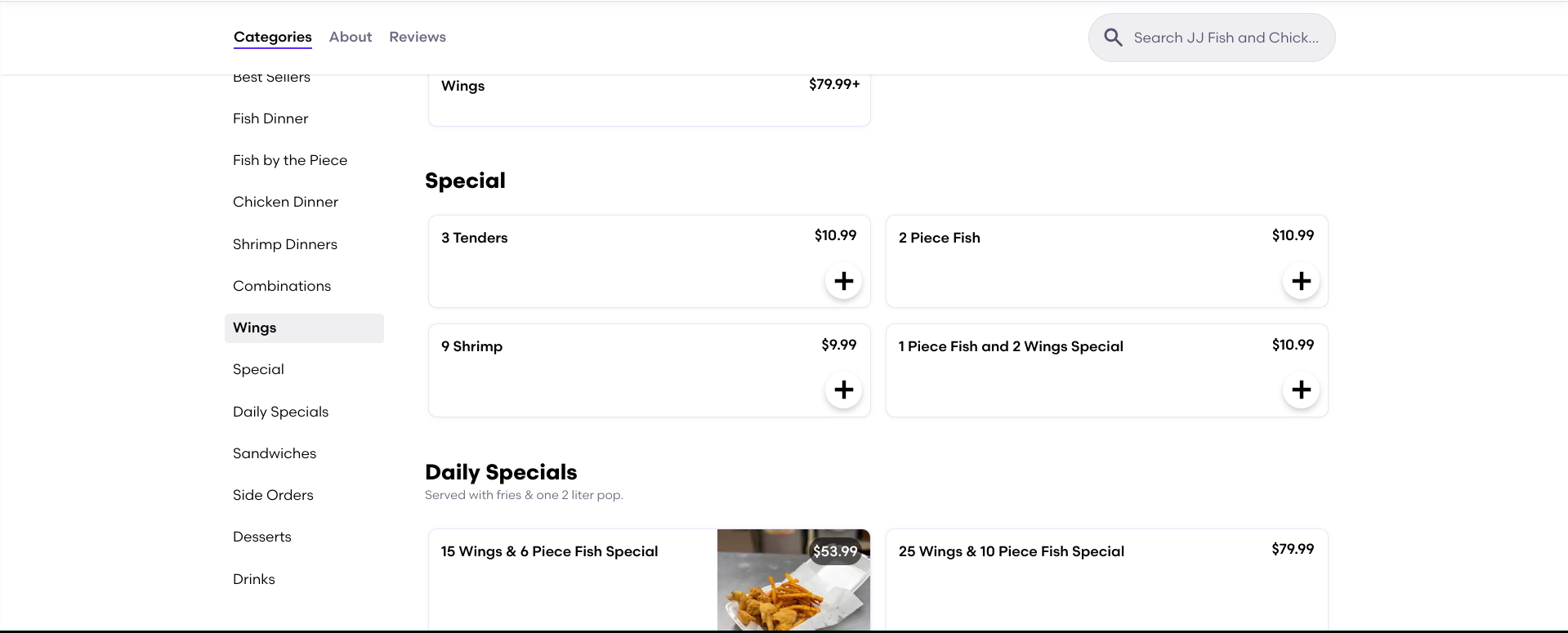}\\
        \begin{minipage}{\textwidth}
            \small
            \texttt{
                \\
                \textbf{Question:} The prices of lasagna and Carolina BBQ Oink Sampler together are how much more than the prices of 3 tenders and 2 pieces of fish together?\\
                \textbf{Answer:} \$493.02
            }
        \end{minipage}
        \caption{An example where UI differences, despite serving the same checkout functionality of a shopping cart, can confound models even with Chain-of-Thought prompting.}
        \label{fig:multi_screen_bbq}
    \end{minipage}
    \hspace{1mm}
    \begin{minipage}{0.48\textwidth}
        \centering
        \includegraphics[width=\textwidth]{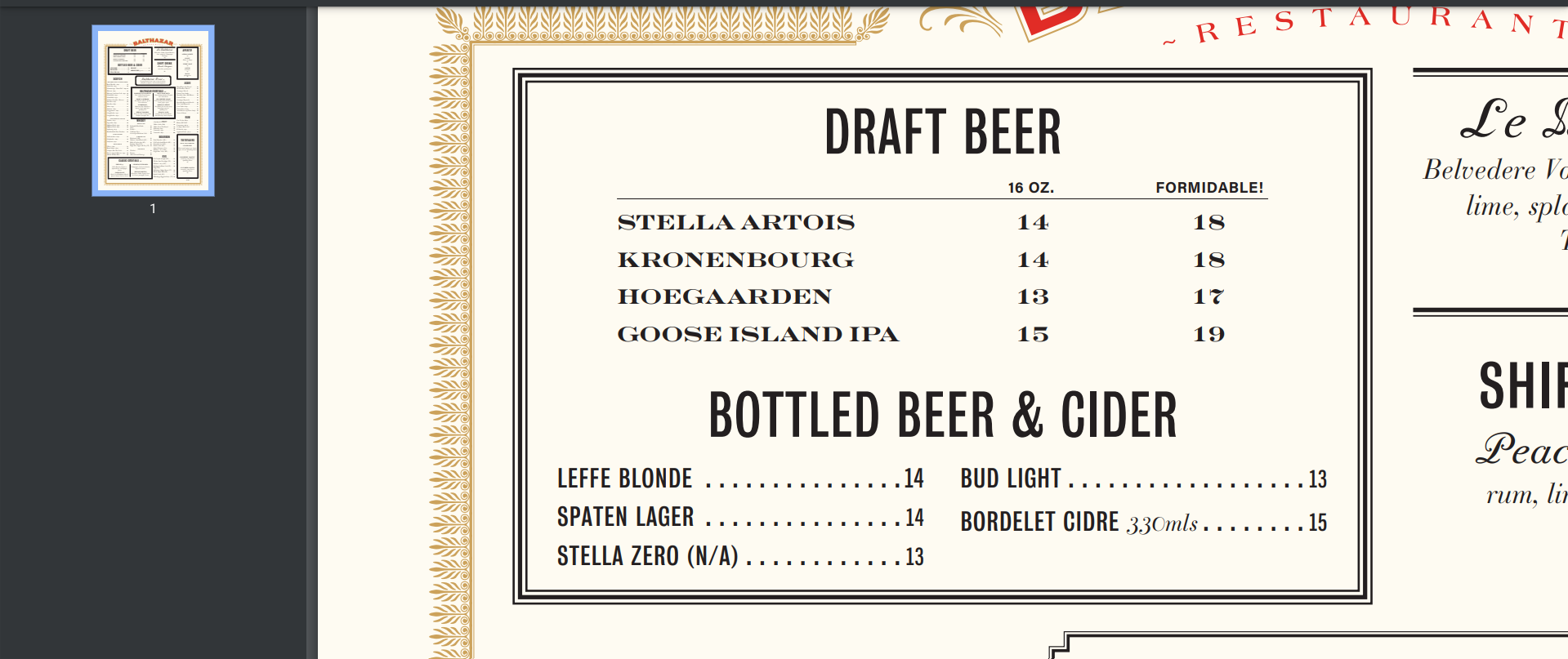}\\
        \includegraphics[width=\textwidth]{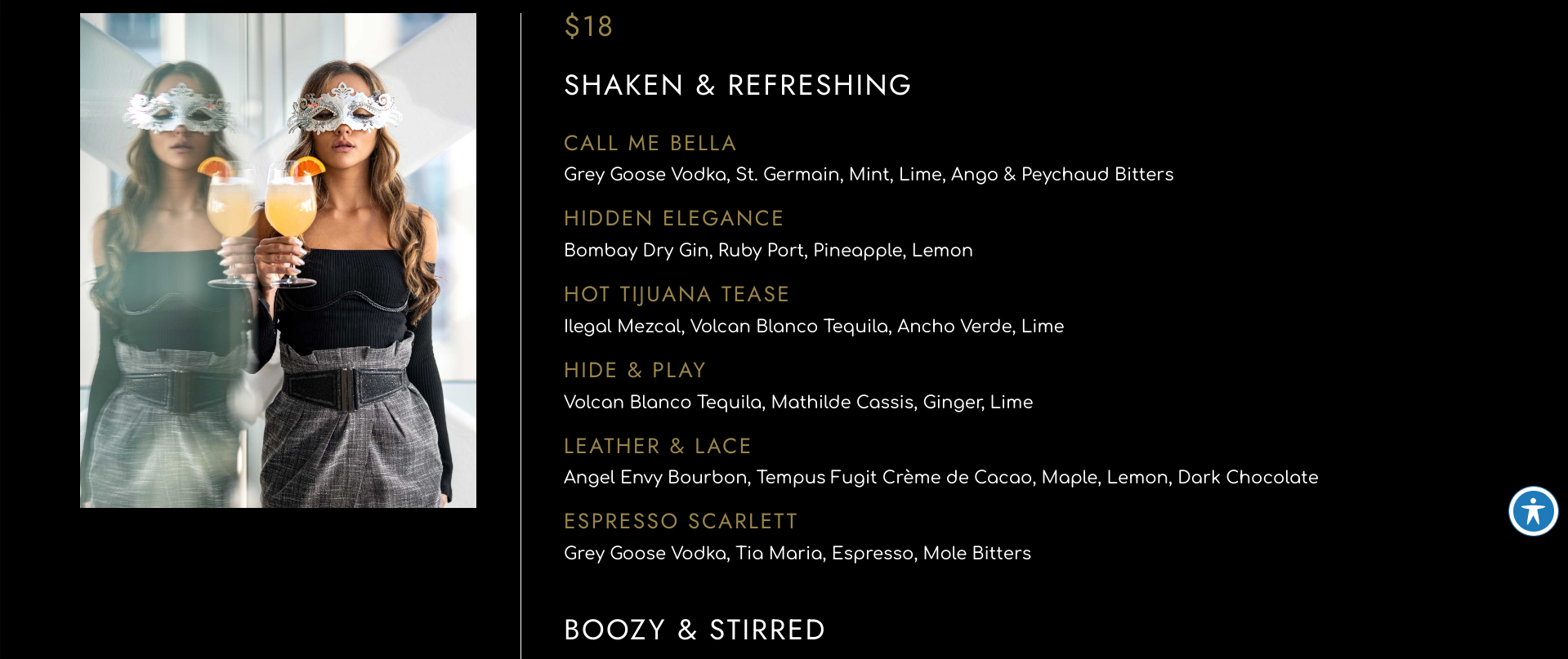}\\
        \includegraphics[width=\textwidth]{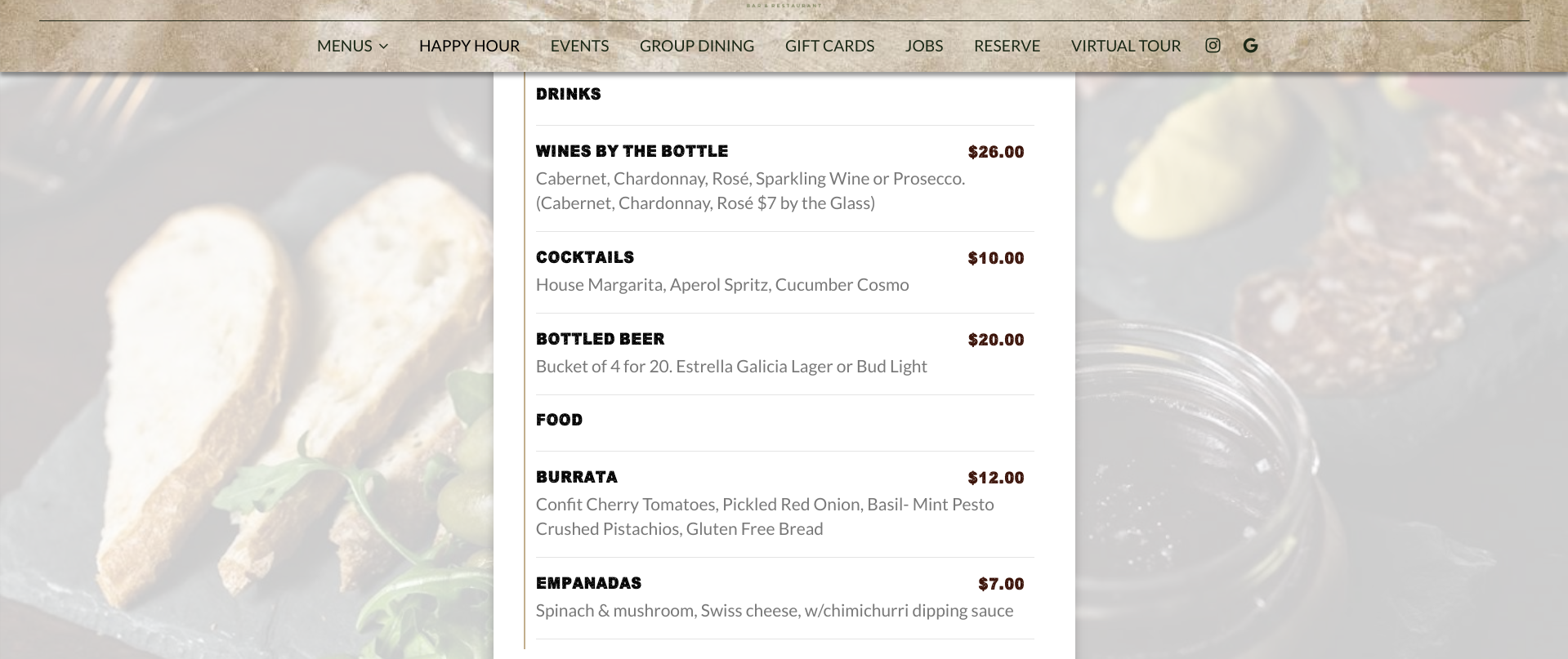}\\
        \begin{minipage}{\textwidth}
            \small
            \texttt{
                \\
                \textbf{Question:} Total, how many drinks are shown on these site pages?\\
                \textbf{Answer:} 18
            }
        \end{minipage}
        \caption{A challenging example that all models failed to answer correctly. The counting task is complicated by the stylistic diversity of the menu. Each models' Chain-of-thought gets it wrong at different steps.}
        \label{fig:multi_screen_drinks}
    \end{minipage}
\end{figure}

The wine example in Figure~\ref{fig:multi_screen_drinks} illustrates some interesting differences between even the same family of models. Claude-3 Sonnet~\cite{anthropic2024claude3} does not recognize the food in page 1 while Claude-3 Opus~\cite{anthropic2024claude3} and Gemini Flash does, suggesting they leverage strong common sense knowledge. However, all models fail to recognize the color and font difference of the menu item subheadings, which are attributes of the drink selections and should not be counted. Gemini Flash was poorer at counting, while understanding the task better. GPT-4V~\cite{openai2024gpt4} tends to hallucinate and make the incorrect associations in its reasoning attributed to reading error (logical fallacy), i.e. mistaking price as number of options and missing the wine option on page 3.

\begin{figure}[h!]
    \centering
    \includegraphics[width=.7\textwidth]{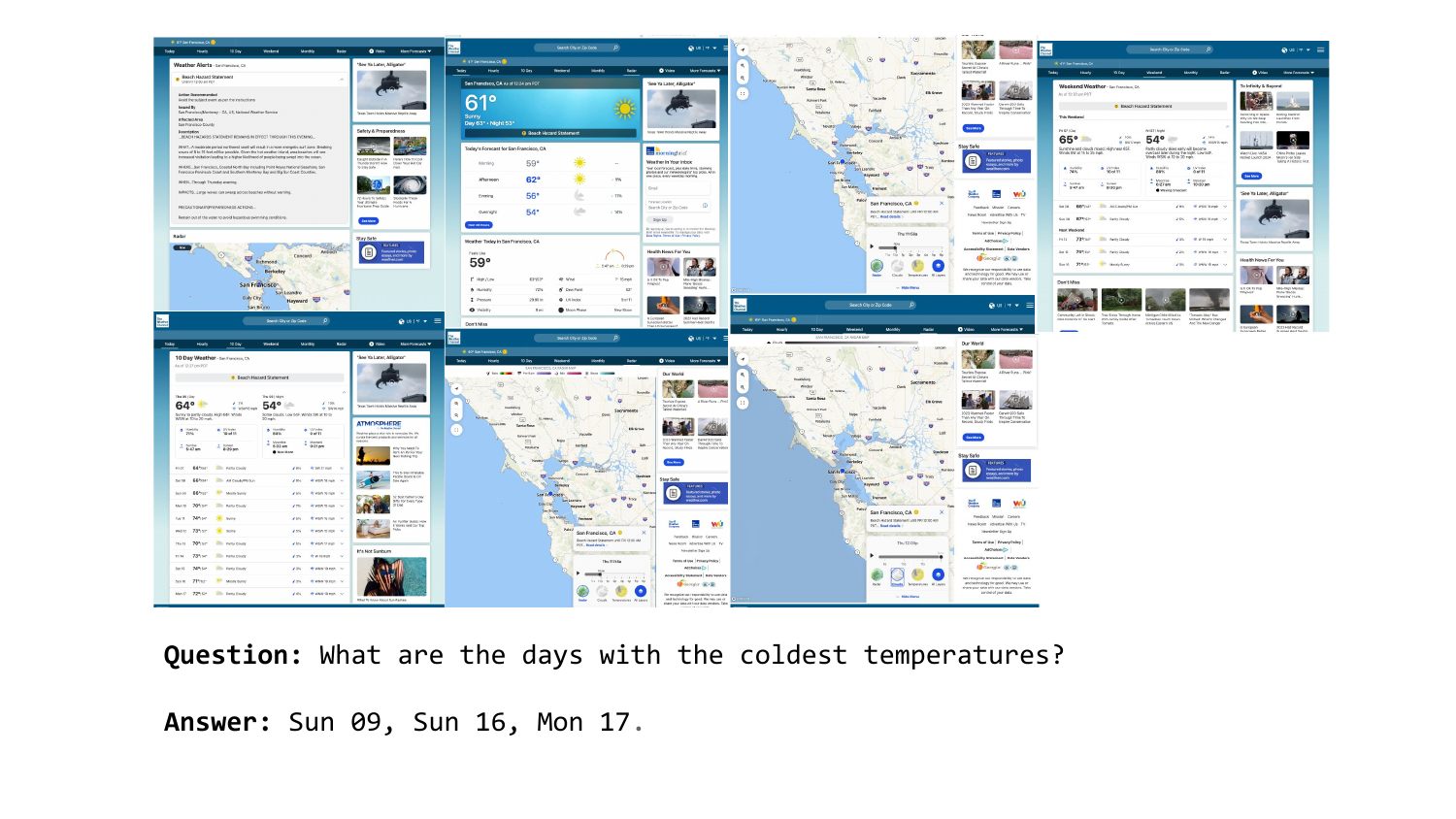}\\
    \caption{Models fails to list correctly the dates with the lowest temperature. This involves first associating each date with its corresponding temperature and then going through each one to decide which one matches the criteria. }
    \label{fig:trace_fail_example}
\end{figure}

\subsubsection{Common errors made by proprietary models}
\begin{itemize}
\item \textbf{Analysis of UI Interfaces}
We noticed all proprietary models tend to be confused by visual differences in UI interfaces serving the same purpose. As displayed in Figure ~\ref{fig:multi_screen_bbq}, when presented with multiple checkout screens, one displaying itemized prices and the other one only the total price, even when GPT-4V misread the individualized price as total price, an example of \textit{visual oversight}.
\item \textbf{Counting instances under a given criteria} Another common failure mode is counting instances under a given criteria or association. These questions often involve counting and filtering based on both semantic and visual attributes, such as dining options, costs, colors and style. (See Figure~\ref{fig:trace_fail_example} for an example.) We found Chain-of-thought step hints to be particularly helpful on decoupling conflated errors across information extraction and reasoning, providing tremendous value in grounding the model's errors and comparing performance between models.  An example is illustrated at Figure ~\ref{fig:multi_screen_drinks}. 
\item \textbf{Considering questions non answerable}
We also observe many errors to be "the question is not answerable" across proprietary models. In these cases, we observe that the main reason is that the model fails to identify relevant information from the screens
\end{itemize}

\subsubsection{Varying performance on different question categories}
For each of the questions in the dataset, we also provide information on the kind of reasoning needed to answer the question. Inspired by similar annotations in the InfographicVQA~\cite{mathew2022infographicvqa} dataset, we use the categories: \emph{arithmetic}, \emph{comparison}, \emph{counting} and \emph{filtering} where each question can have more than one category. We use Gemini 1.5 Pro~\cite{team2023gemini} to optionally assign one or more of the categories to each question. We manually inspect all the categories assigned and make necessary corrections. As shown in figure~\ref{fig:results-question-type}, we note that different models exhibit different performances across the various categories. Across all models, questions related to counting tend to get a lower accuracy than other ones, while questions involving only arithmetic operations get a higher score. However, it is important to note that these aggregate numbers relying on the final answer are conflating errors due to reasoning with errors due to information extraction or other causes. We also note that the performance decreases as the number of entities across the screens increases indicating the model may not able to extract all the information correctly when a large number of entities are present.

\subsubsection{Open source model performance}
We observe that, open source models perform worse than proprietary LLMs on WebQuest tasks. We believe that the difference in the sizes of the open source models compared to proprietary LLMs is a significant factor here. We observed that for BLIP-2 variants ~\cite{li2023blip2bootstrappinglanguageimagepretraining,dai2023instructblip}, Chain-of-thought prompting~\cite{wei2022chain} did not result in a more detailed output and hence show no performance improvement compared to prompts with no CoT. We also note that InstructBLIP~\cite{dai2023instructblip} was trained with an image resolution of 224x224. However, WebQuest has screenshots with significantly higher resolutions and requires the understanding of the fine grained details of text and UI elements present on the screen. Unsuprisingly, we noticed performance improvement with inference at a resolution of 448x448 but did not see further improvements with increasing image resolution.

\section{Conclusion}
\label{sec:conclusions}
In this work, we introduce WebQuest, the first multimodal question answering benchmark designed to evaluate reasoning abilities across multiple screens. Our benchmark encompasses three distinct settings: single-screen, multi-screen, and trace-based, enabling a comprehensive assessment of model capabilities.  Evaluations of state-of-the-art MLLMs, including Gemini Flash \cite{geminiteam2024gemini}, Claude 3 \cite{anthropic2024claude3}, and GPT-4V \cite{openai2024gpt4}, reveal a significant performance gap between single-screen and multi-screen reasoning tasks. Notably, Chain-of-thought prompting \cite{wei2022chain} proves effective for navigating complex, multi-screen scenarios by facilitating information extraction and synthesis across multiple screens.
Future research directions include leveraging richer screen information, such as OCR, DOM, and screen annotations from tools like ScreenAI \cite{baechler2024screenai}, to potentially enhance model performance and even replace raw screen images in cases of long screen sequences. Furthermore, extending WebQuest to support personalized, multi-turn dialogue scenarios presents a compelling avenue for future exploration, enabling the development of agents capable of assisting users in dynamic, context-aware conversations. \todo{sri: Maybe add a line about memory or retrieval. Mention this as a potential future research direction.}

\section*{Limitations and Risks}
\label{sec:limitations}

Some of the limitations of our work which we plan to address in future research are: 

\begin{itemize}
    \item 
    Our current dataset does not include multi-turn interactions which contain follow up questions or clarifications to previous questions.
    \item
    To avoid any personally identifiable information in the screen data, we do not include screens that require user specific credentials.
    \item
    The question and answers, navigation traces collected may reflect cultural and demographic biases of the raters.
    
\end{itemize}

\clearpage
\bibliographystyle{ACM-Reference-Format}
\bibliography{bibliography}

\clearpage

\begin{appendices}
    \section*{APPENDIX}
    \section{Dataset Examples}
\label{app:examples}

\begin{figure}[pt]
    \centering
    \begin{subfigure}[t]{0.48\textwidth}
        \includegraphics[width=\textwidth]{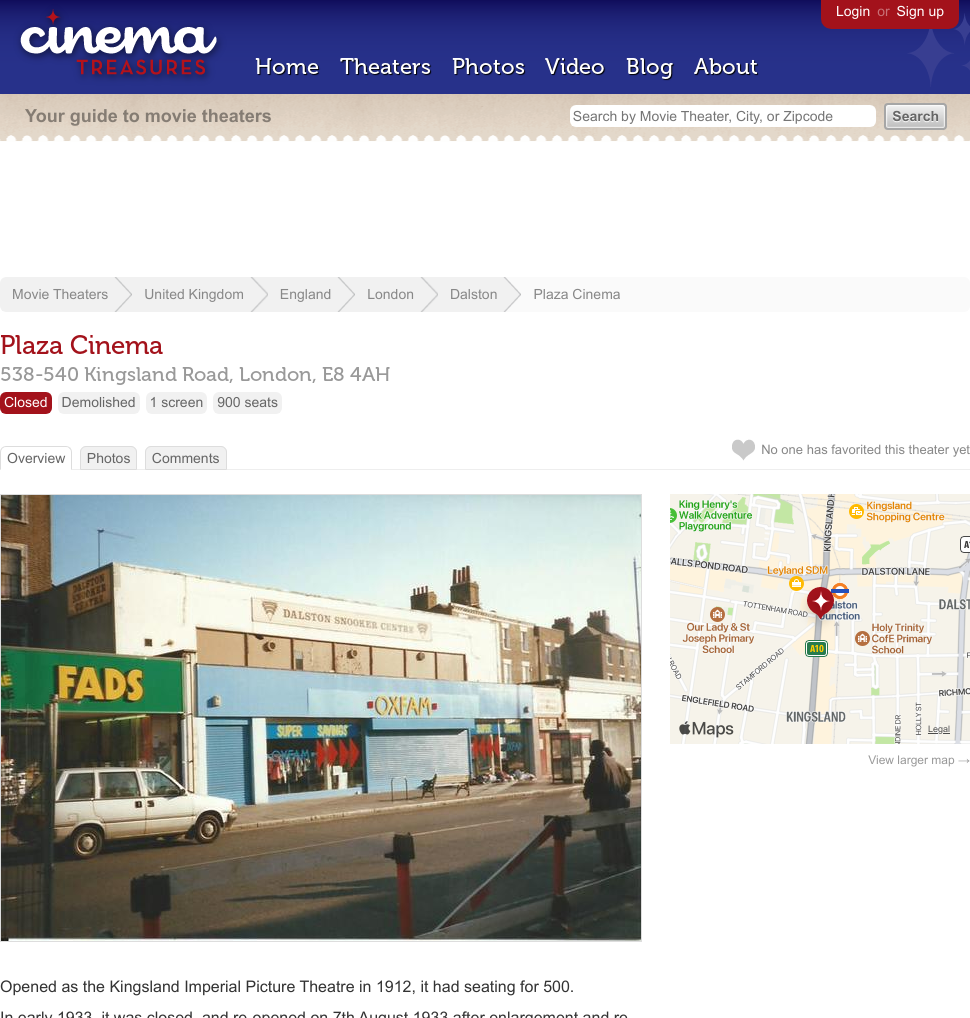}\\
        \begin{minipage}[b]{\textwidth}
            \small
            \texttt{
                \\
                \textbf{Question:} How many more seats does the Plaza Cinema have than it had when it opened?\\
                \textbf{Answer:} 400
            }
        \end{minipage}
        \subcaption{The task is to retrieve and compute the difference between the number of seats mentioned in different context of the screen layout.}
        \label{fig:single_page_cinima_seats}
    \end{subfigure} \hfill
    \begin{subfigure}[t]{0.48\textwidth}
        \includegraphics[width=\textwidth]{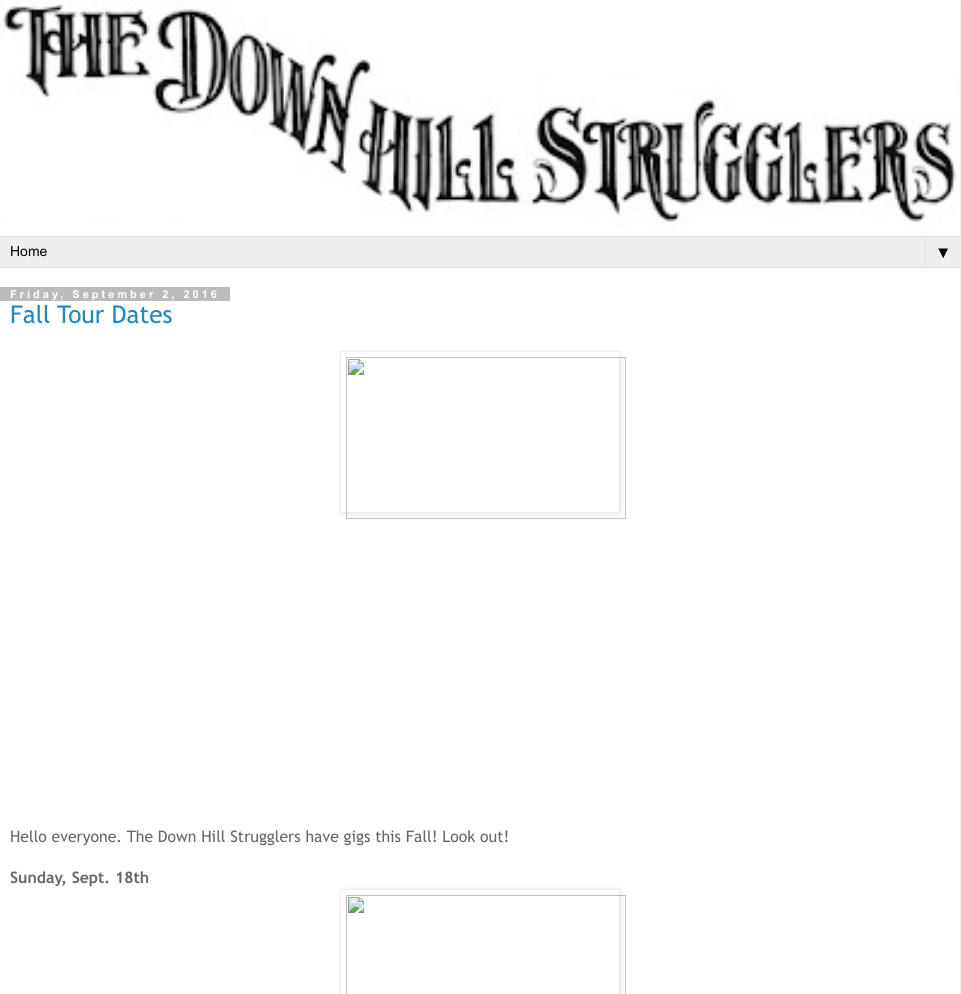}
        \begin{minipage}[b]{\textwidth}
            \small
            \texttt{
                \\
                \textbf{Question:} How many days are there between the two gigs?\\
                \textbf{Answer:} 16
            }
        \end{minipage}
        \subcaption{The task is to retrieve 2 dates and compute the difference between them in days. Note that the dates are in different colors of font and background, and in different date format. Also note the web page display show two images are missing, which is a common case the QA task is able to handle. We intentionally include screenshots as such.}
        \label{fig:single_page_two_gigs}
    \end{subfigure} \hfill
    \begin{subfigure}[t]{0.48\textwidth}
        \includegraphics[width=\textwidth]{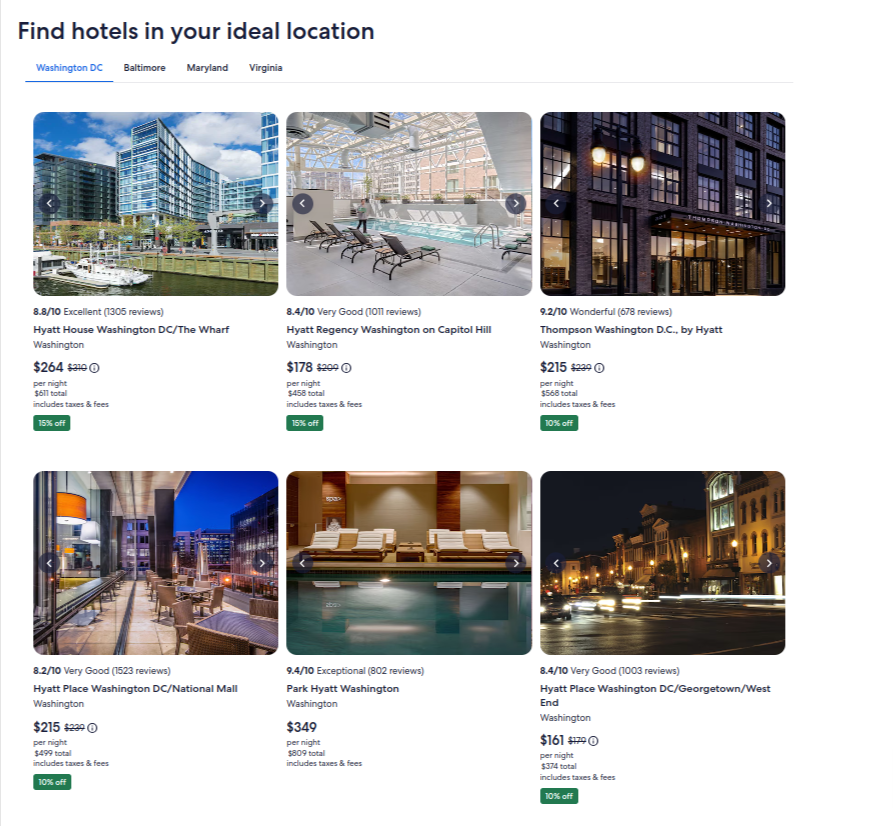}
        \begin{minipage}[b]{\textwidth}
            \small
            \texttt{
                \\
                \textbf{Question:} What is the maximum discount?\\
                \textbf{Answer:} 15\% off
            }
        \end{minipage}
        \subcaption{The task is retrieve all the discounts on the screen and compare them against each other.}
        \label{fig:single_page_two_gigs}
    \end{subfigure}
    \caption{Examples of Single Screen QA.}
    \label{fig:single_page_examples}
\end{figure}

\begin{figure}[pt]
    \centering
    \includegraphics[width=.48\textwidth]{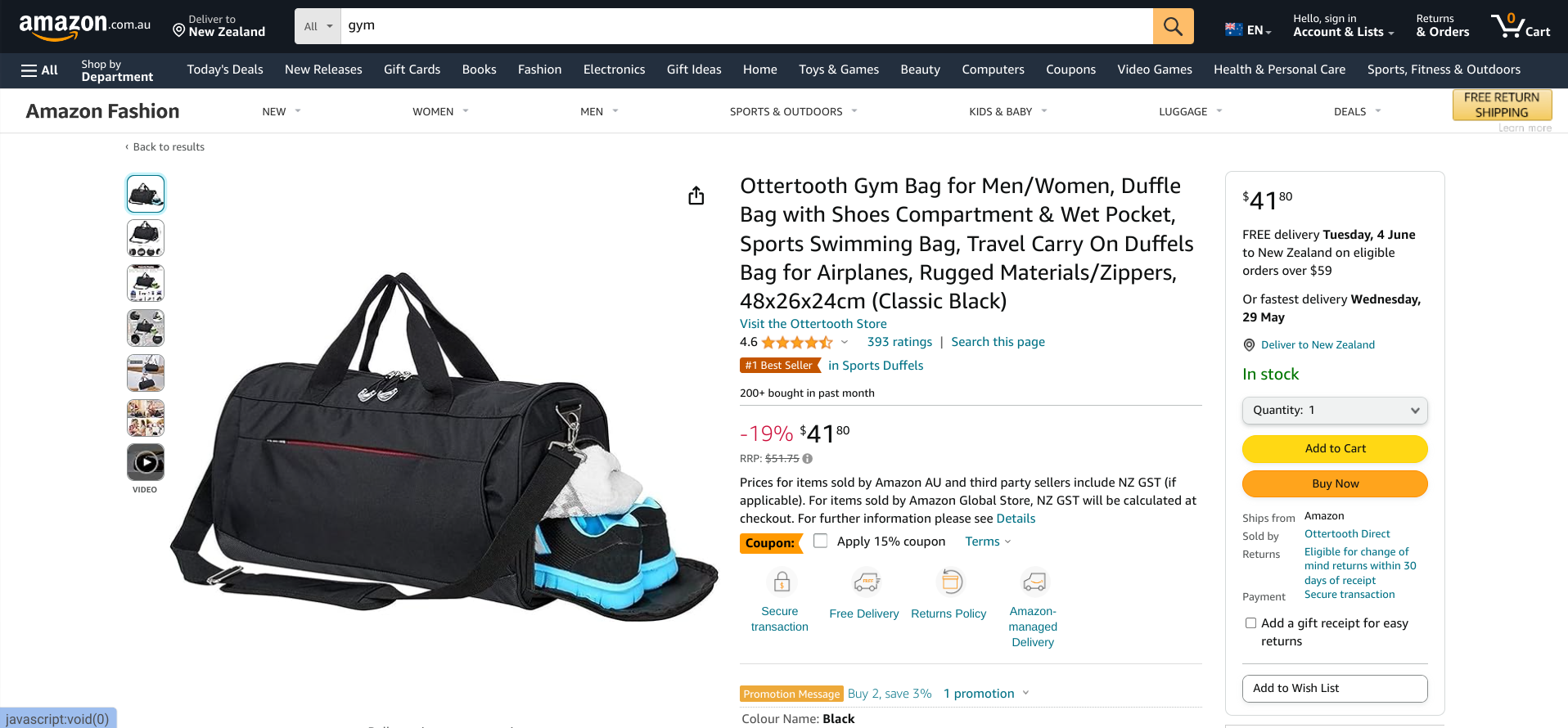}\\
    \includegraphics[width=.48\textwidth]{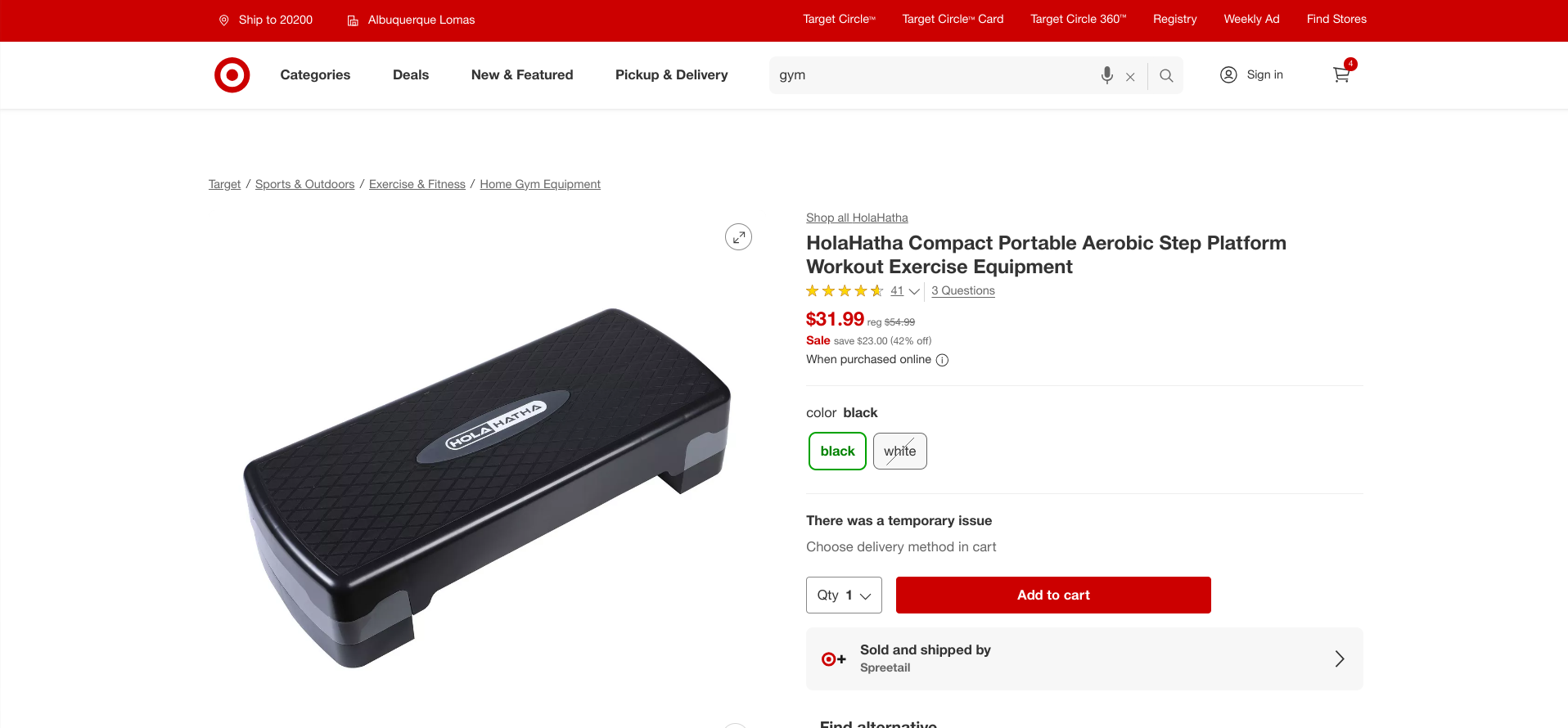}\\
    \includegraphics[width=.48\textwidth]{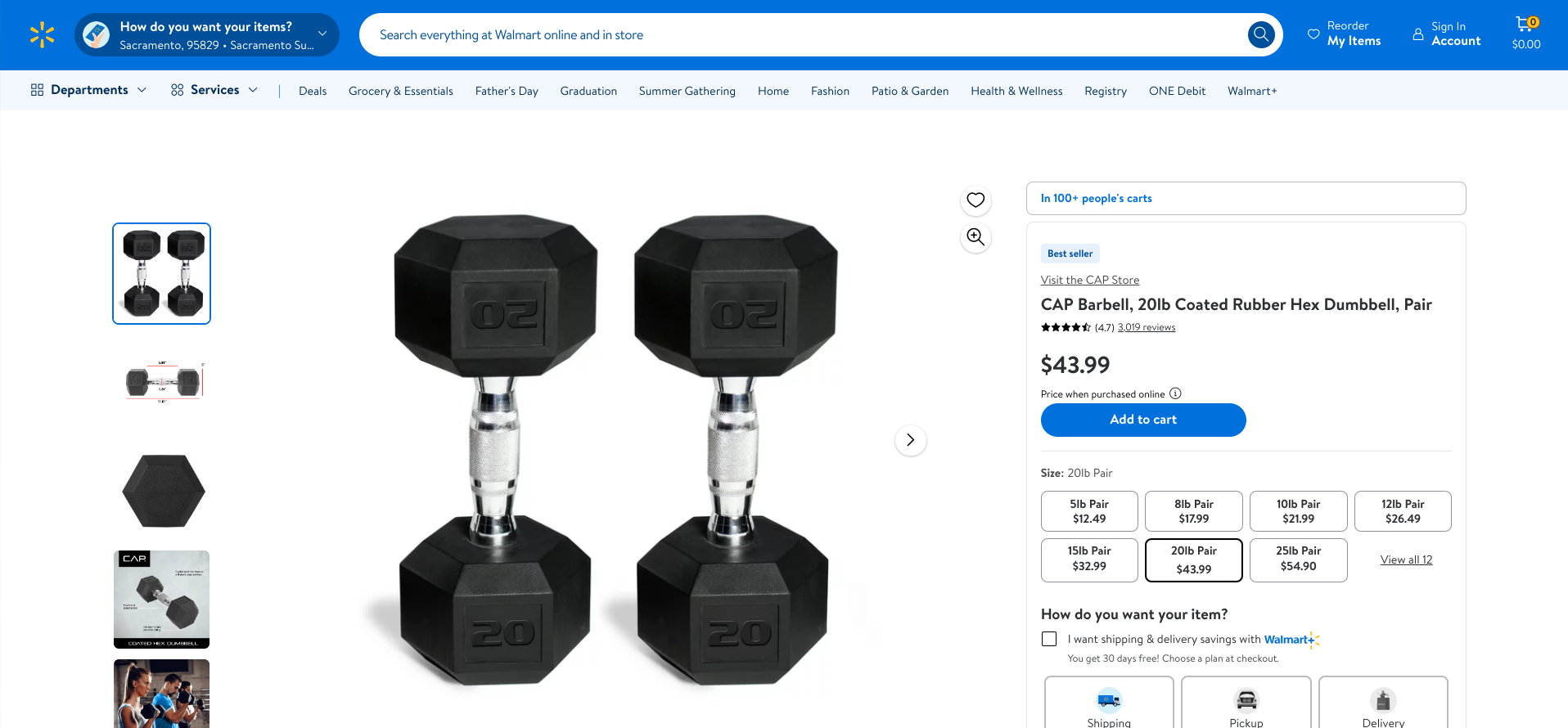}\\
    \includegraphics[width=.48\textwidth]{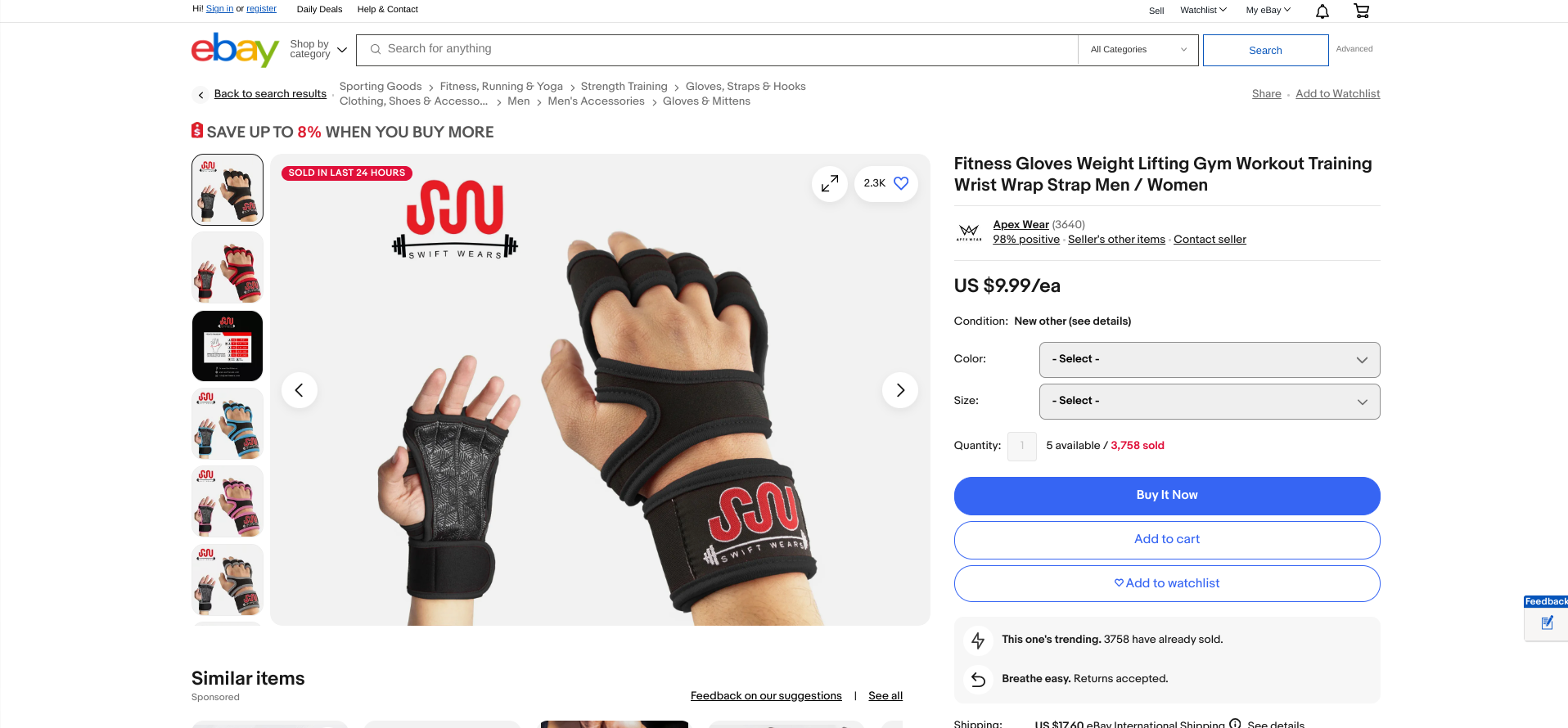}\\
    \begin{minipage}{.48\textwidth}
        \small
        \texttt{
            \\
            \textbf{Question:} How much do all the gym accessories cost in total?\\
            \textbf{Answer:} \$127.77
        }
    \end{minipage}
    \caption{An example of Multipage QA, about prices of selected gym accessories, each from a different online shop. Notice the layouts and font sizes vary among them.}
    \label{fig:multi_page_gym_accessories_cost}
\end{figure}

\begin{figure}
    \centering
    \includegraphics[width=.48\textwidth]{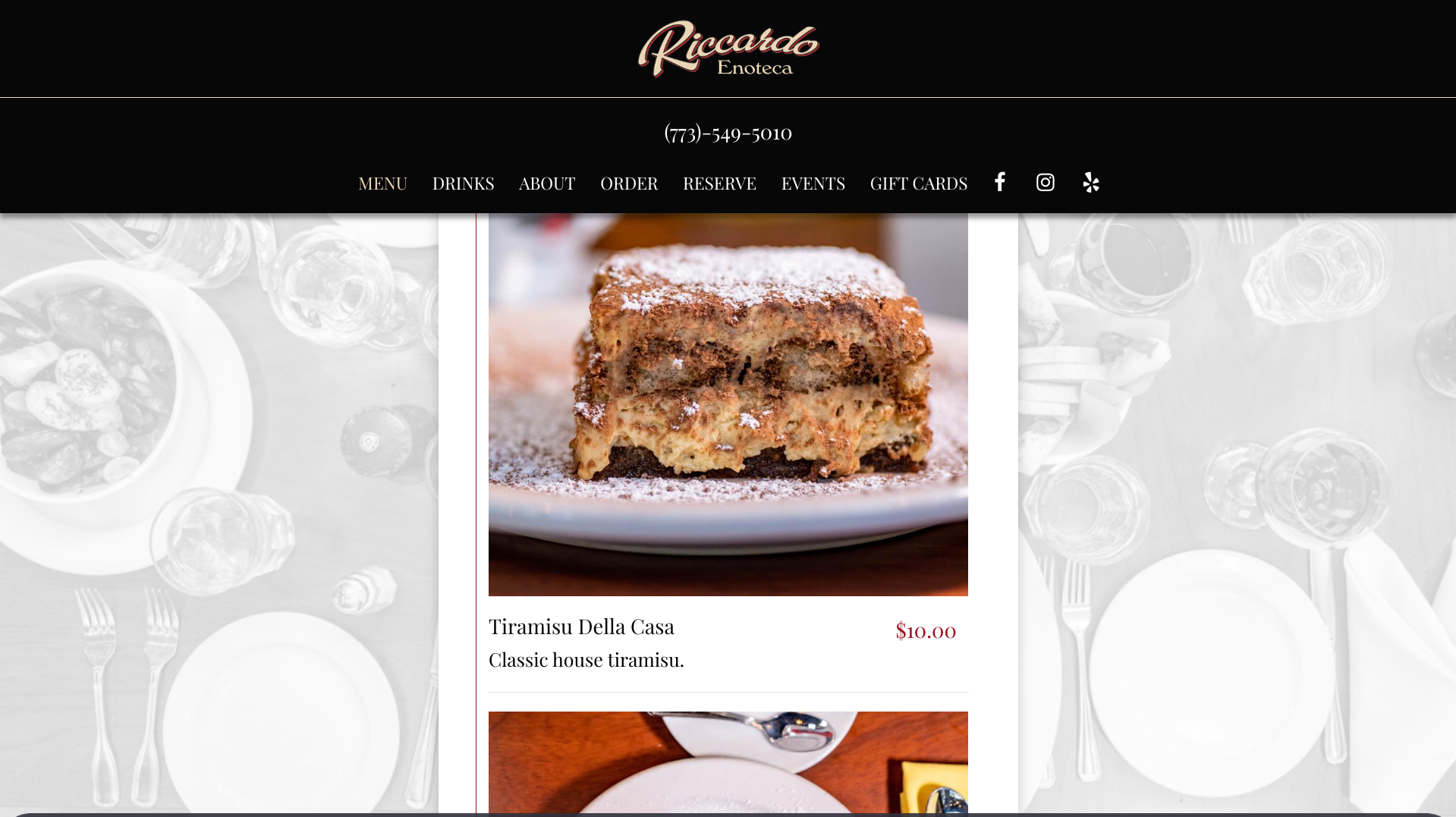}\\
    \includegraphics[width=.48\textwidth]{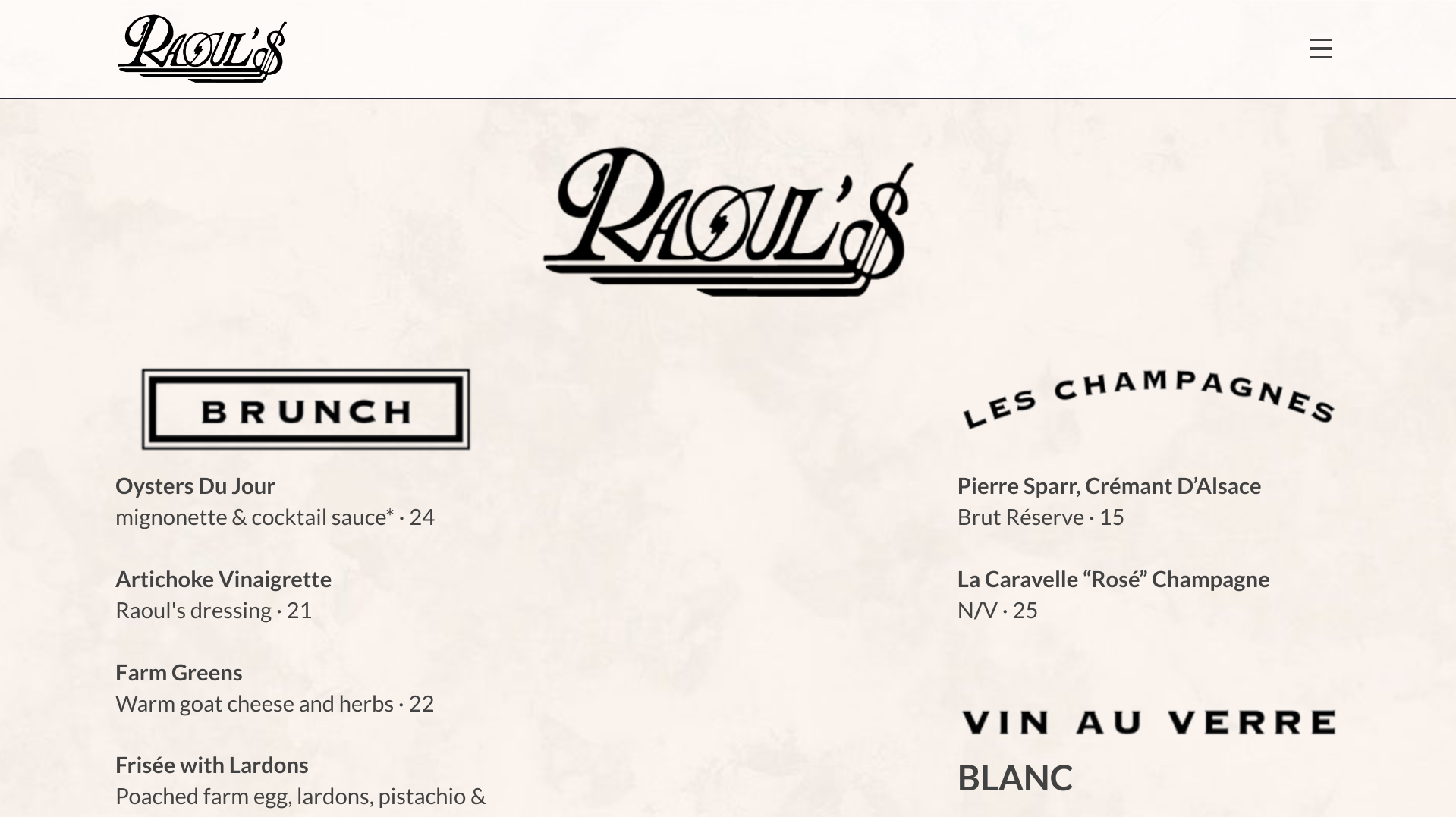}\\
    \includegraphics[width=.48\textwidth]{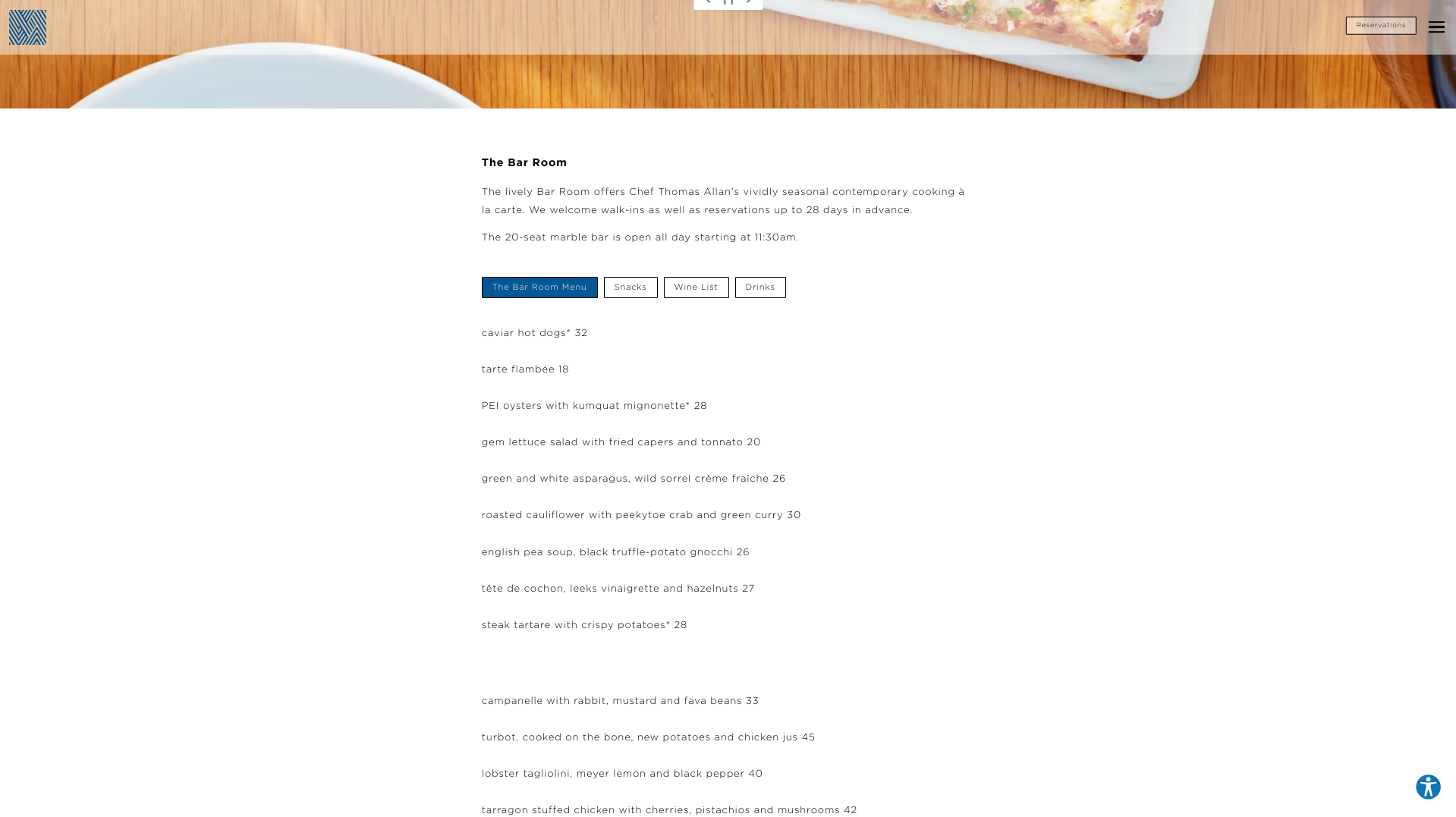}\\
    \includegraphics[width=.48\textwidth]{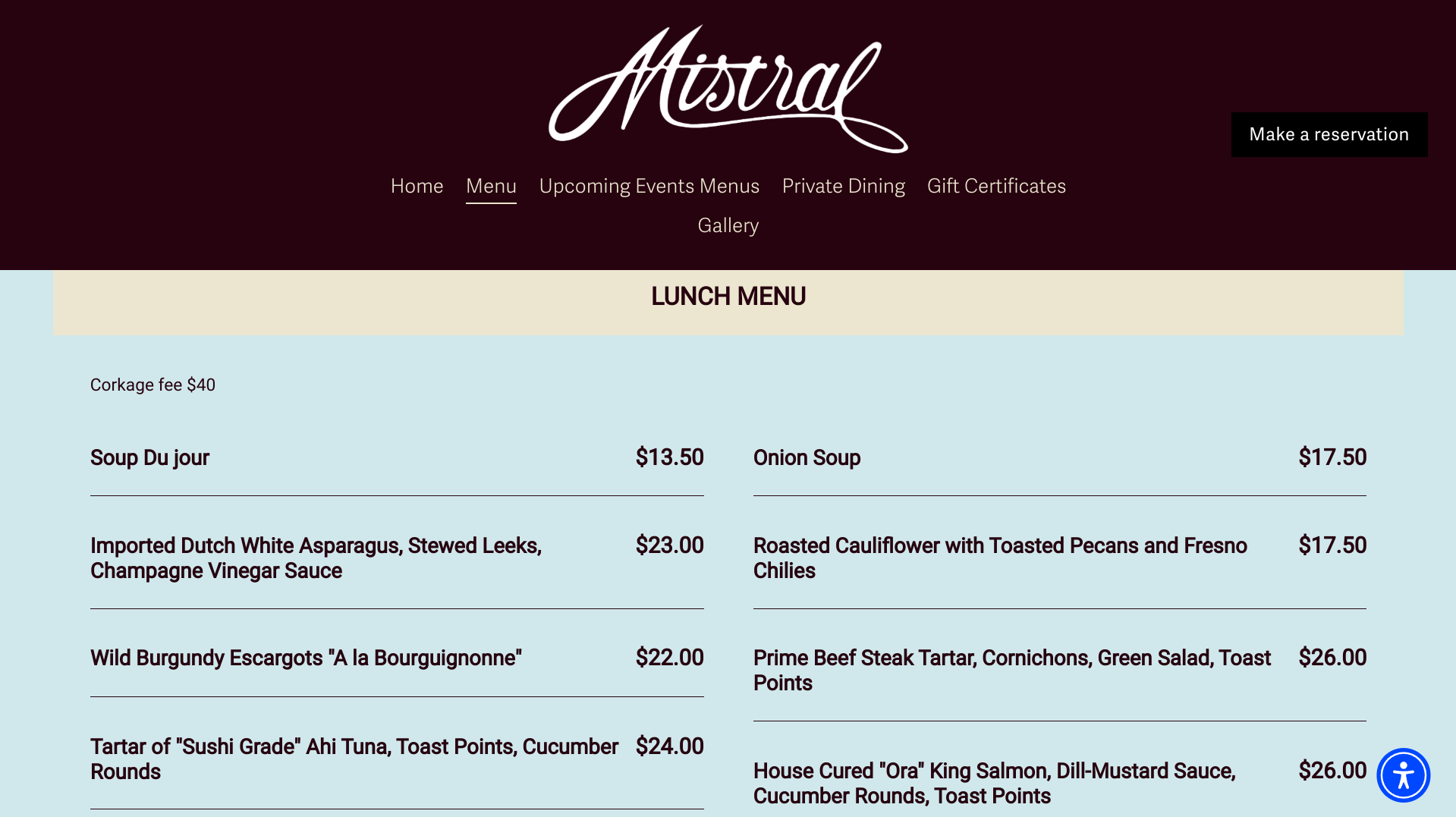}\\
    \begin{minipage}{.48\textwidth}
        \small
        \texttt{
            \\
            \textbf{Question:} The sum of the prices of soup do jour, caviar hot dogs, and oysters do jour is how much more than the price of Tiramisu Della Casa?\\
            \textbf{Answer:} \$59.5
        }
    \end{minipage}
    \caption{An example of Multipage QA, about prices of food items over 4 websites of restaurant menus. Notice the layouts and font sizes vary significantly among them.}
    \label{fig:multi_page_food_price_diff}
\end{figure}

\begin{figure*}[h]
    \centering
    \begin{tabular}{ccc}
        \includegraphics[width=.3\textwidth]{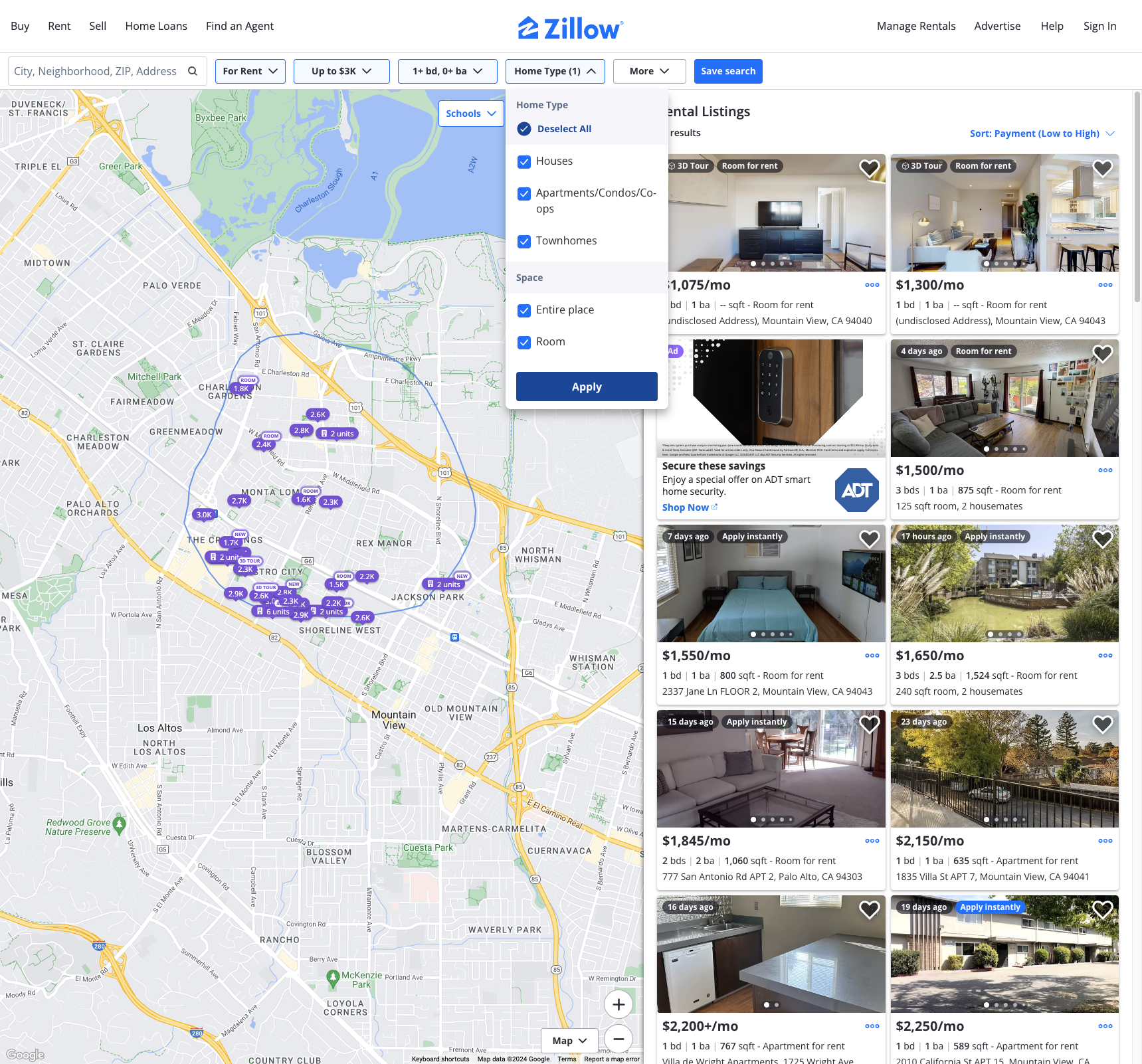} &  \includegraphics[width=.3\textwidth]{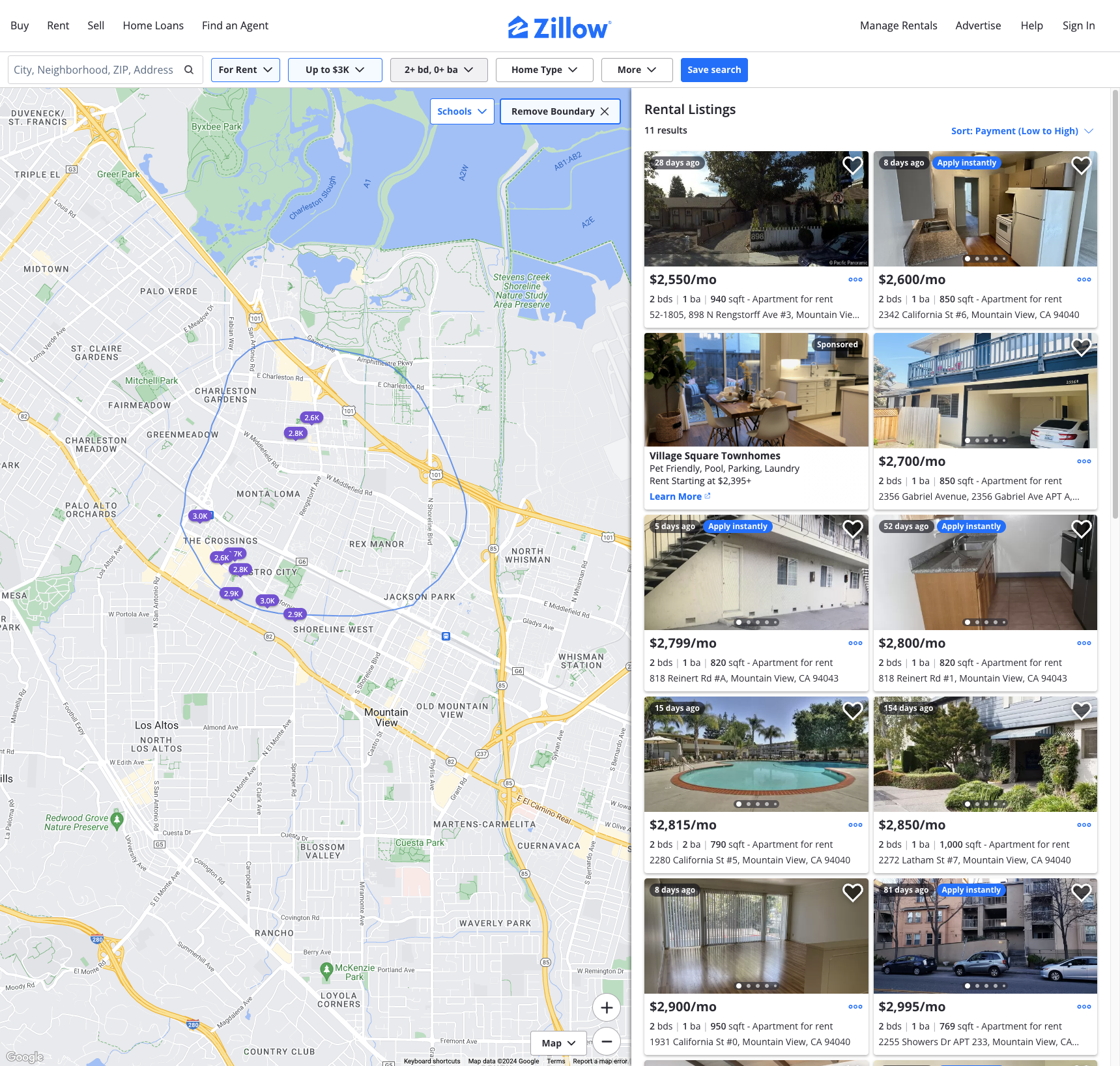} & \includegraphics[width=.3\textwidth]{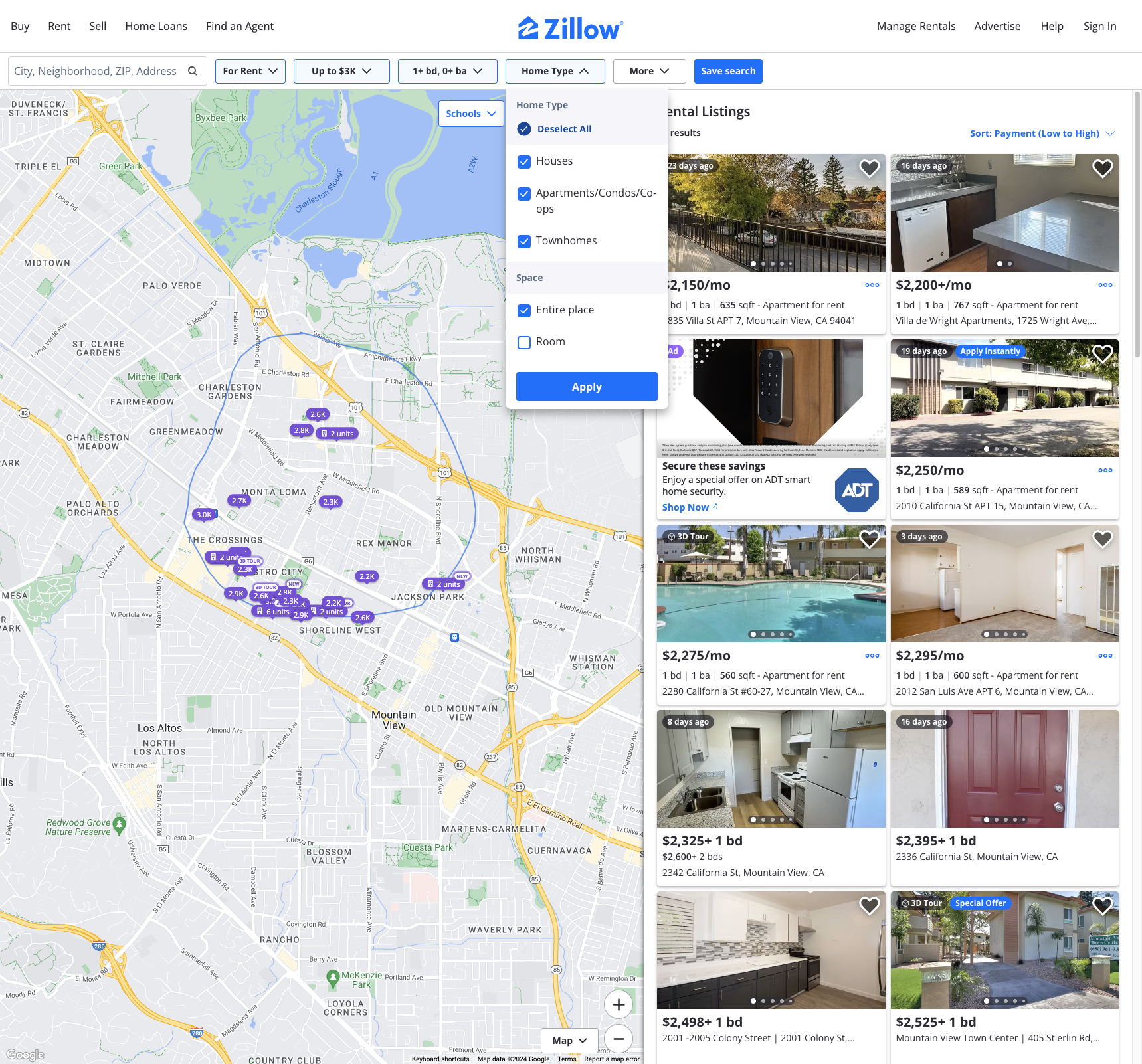} \\
        \includegraphics[width=.3\textwidth]{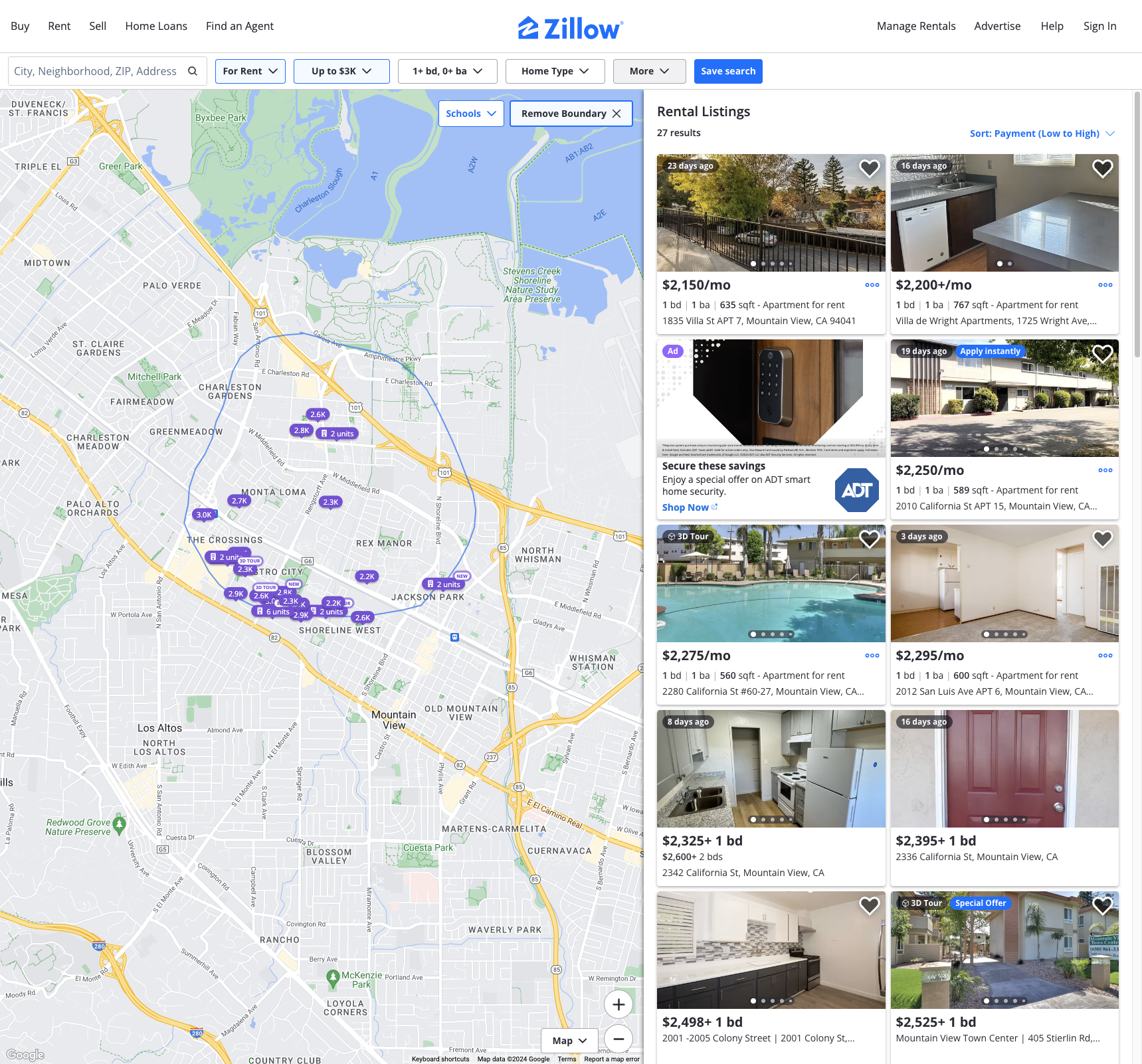} &  \includegraphics[width=.3\textwidth]{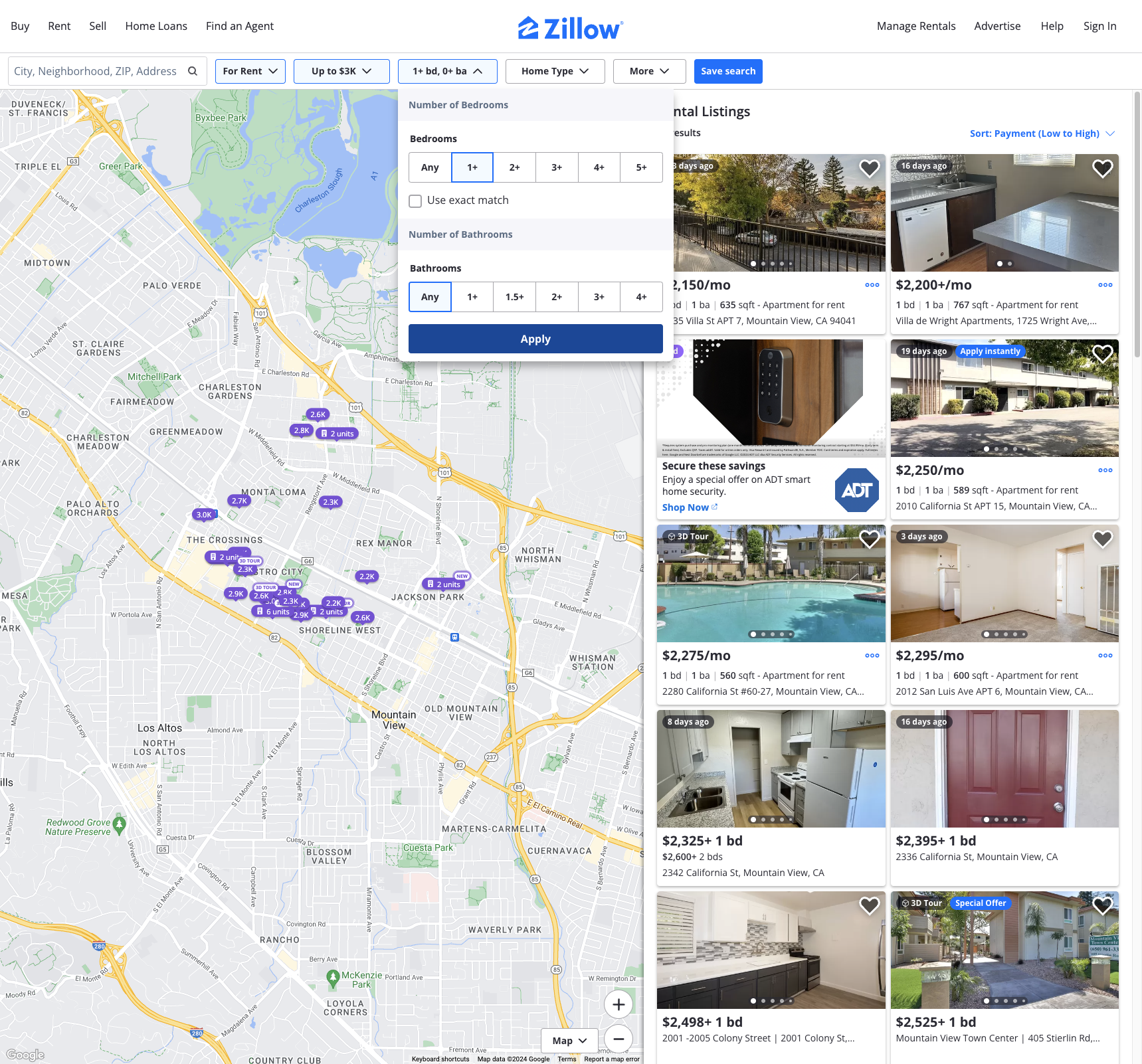} & \includegraphics[width=.3\textwidth]{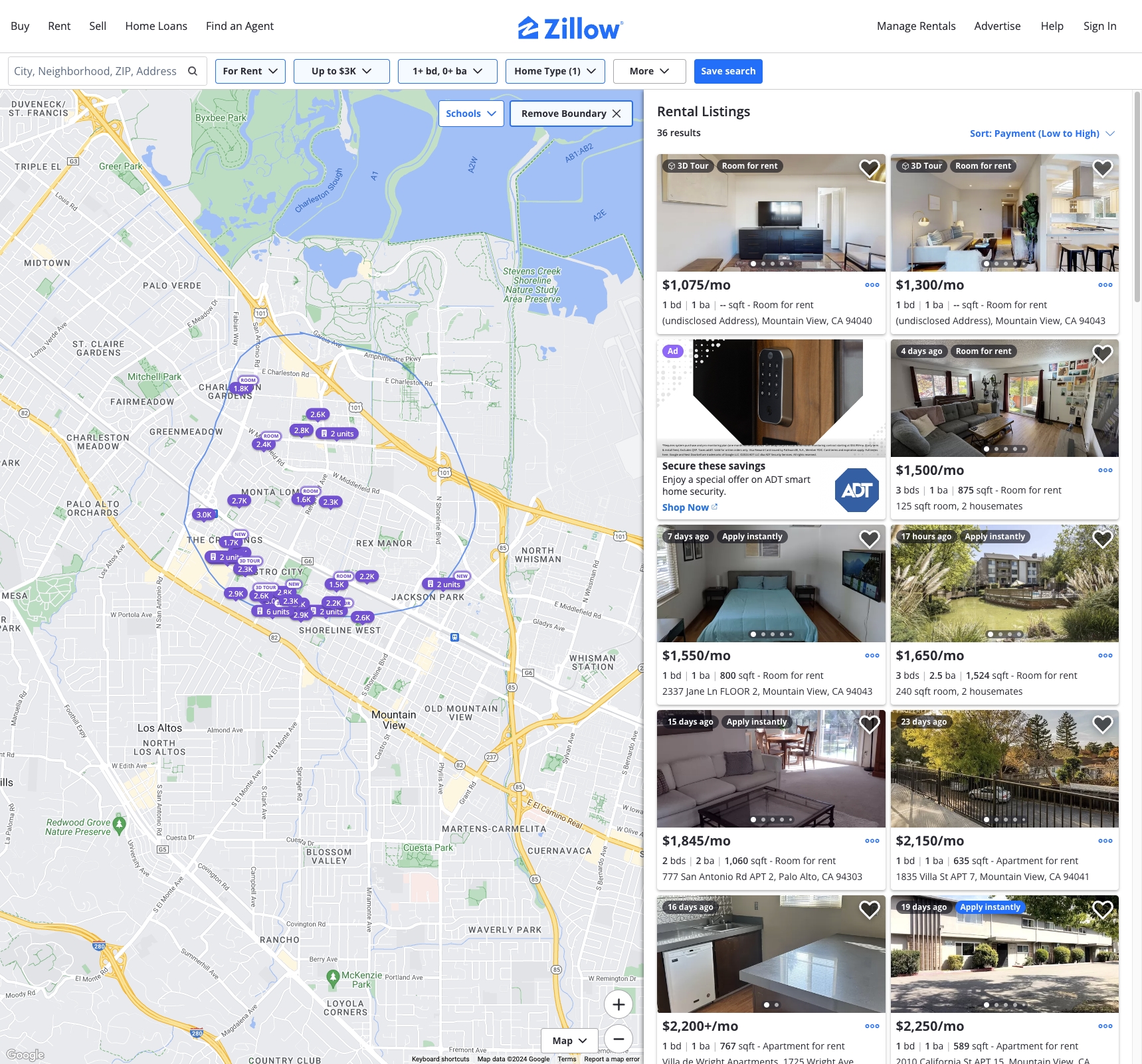} \\
        \includegraphics[width=.3\textwidth]{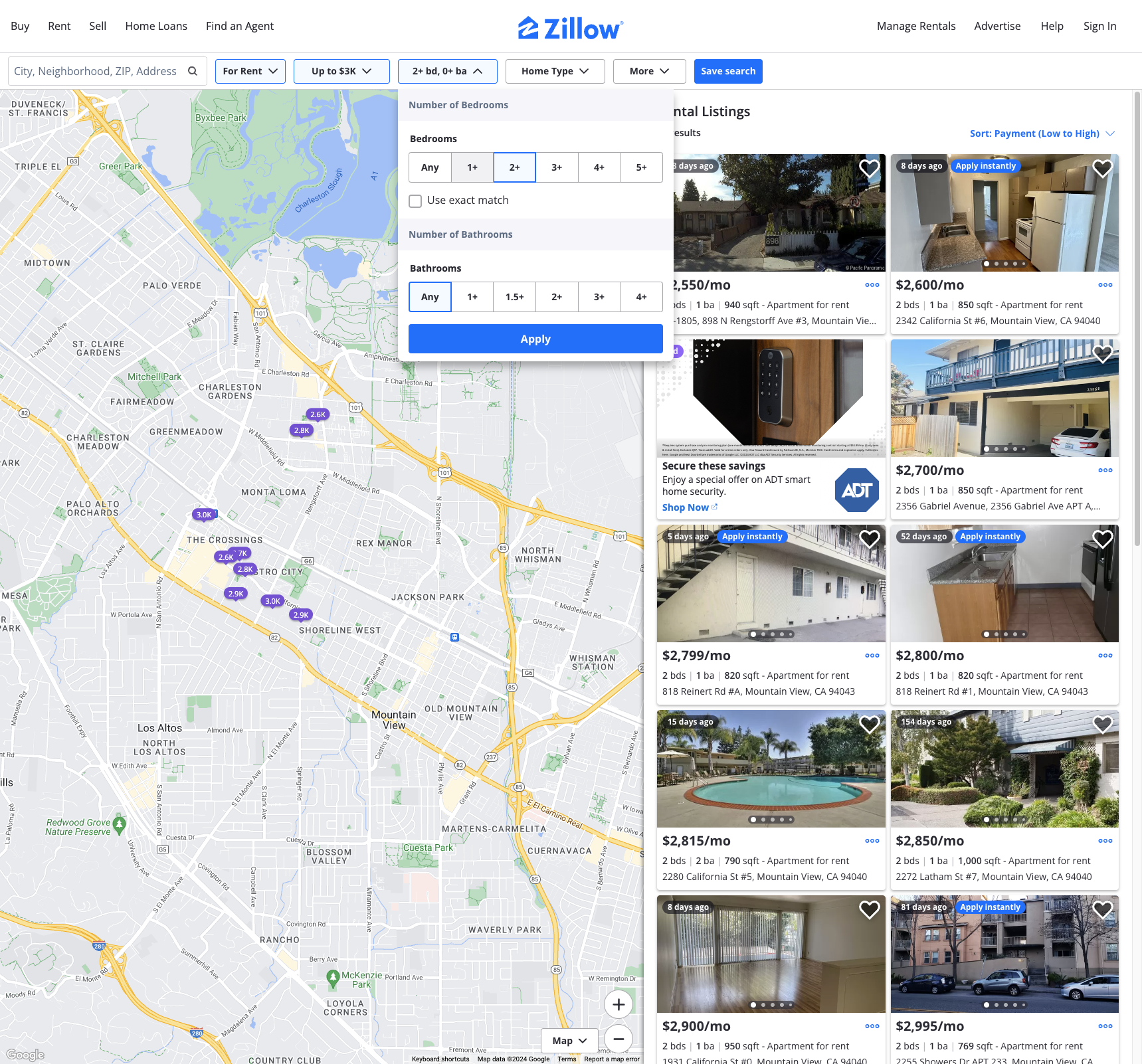} &  \includegraphics[width=.3\textwidth]{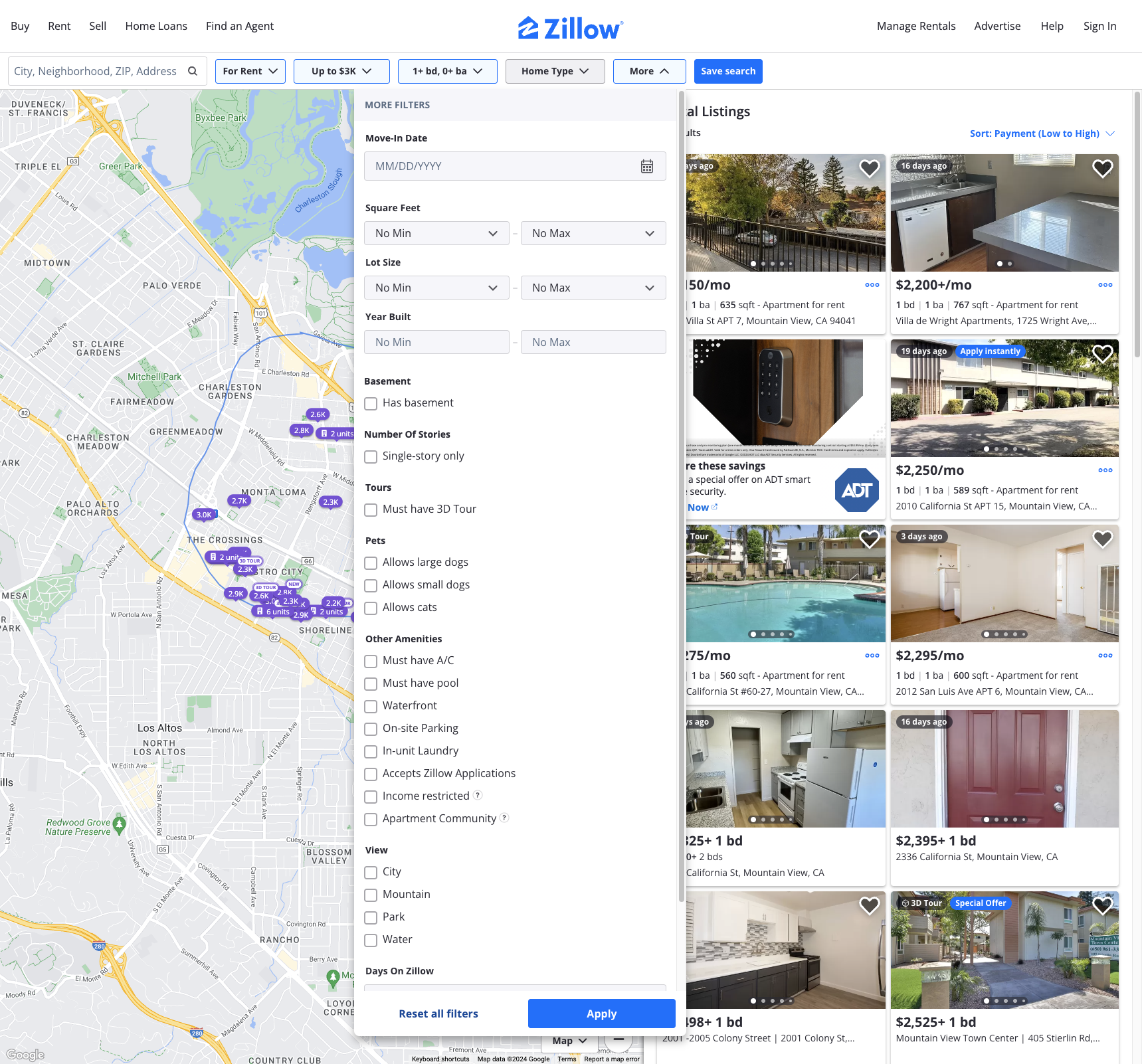} &
        \begin{minipage}[b]{.3\textwidth}
            \small
            \texttt{
                \\
                \textbf{Question:} To rent for an entire place, how much I can save per month if I rent a bedroom instead of two, if I rent the cheapest one?\\
                \textbf{Answer:} \$400
                \\
            }
        \end{minipage}
    \end{tabular}
    \caption{An example of Trace QA. The example, situated in a rental exploration scenario, is about price differences between different options. Notice the trace captures the steps taken in the exploration.}
    \label{fig:trace_rental_price_diff}
\end{figure*}

We show a few examples of the WebQuest dataset, in
Figure~\ref{fig:single_page_examples} for Single Screen QA, in Figures~\ref{fig:multi_page_gym_accessories_cost} and~\ref{fig:multi_page_food_price_diff} for Multi Screen QA, and in Figure~\ref{fig:trace_rental_price_diff} for Trace QA.

    \clearpage
    \section{Statistics of the Dataset}

We show detailed statistics of the dataset in
Table~\ref{tab:stats_single_screen}, Table~\ref{tab:stats_multi_screen}, 
Table~\ref{tab:category_multi_screen},
and Table~\ref{tab:stats_trace}.

\begin{table}[H]
    \centering
    \makebox[\textwidth]{ %
        \begin{minipage}[t]{0.45\textwidth} %
            \centering
            \begin{tabular}{|c|c|} 
                \hline
                \textbf{Category} & \textbf{Num Examples} \\ [0.5ex] 
                \hline
                Restaurant & 44 \\ 
                \hline
                Shopping & 184 \\ 
                \hline
                Hotel & 68 \\ 
                \hline
                Transportation & 114 \\ 
                \hline
                Landmark & 21 \\ 
                \hline
                News & 22 \\ 
                \hline
                Media Sharing & 7 \\ 
                \hline
                Education \& Information & 9 \\ 
                \hline
                Movies & 24 \\ 
                \hline
                Real Estate & 19 \\ 
                \hline
                Other & 30 \\ 
                \hline
                \textbf{Total} & \textbf{542} \\ 
                \hline
            \end{tabular}
            \caption{Distribution of the 542 examples of Single Screen QA, per categories and the number of screens in the example.}
            \label{tab:stats_single_screen}
        \end{minipage}
        \hfill
        \begin{minipage}[t]{0.45\textwidth} %
            \centering
            \begin{tabular}{|c|c|} 
                \hline
                \textbf{Category} & \textbf{Num Examples} \\ [0.5ex] 
                \hline
                Shopping & 209  \\ 
                \hline
                Restaurant & 72 \\
                \hline
                Transportation & 12  \\ 
                \hline
                Hotel & 12  \\ 
                \hline
                Landmark & 1 \\
                \hline
                Other & 1 \\
                \hline
                \textbf{Total} & \textbf{307} \\ 
                \hline
            \end{tabular}
            \caption{Distribution of the 307 examples of Multi Screen QA per categories.}
            \label{tab:category_multi_screen}
        \end{minipage}
    }

\vspace{2pt}

    \centering
    \begin{tabular}{|c|c|c|c|} 
        \hline
        \textbf{3 screens} & \textbf{4 screens} & \textbf{5 screens} & \textbf{Total} \\
        \hline
        54 & 131 & 122 & \textbf{307}\\ 
        \hline
    \end{tabular}
    \caption{Distribution of the 307 examples of Multi Screen QA per the number of screens in the example.}
    \label{tab:stats_multi_screen}

\vspace{5pt}

\vspace{5pt}

    \begin{tabular}{|c|c|} 
        \hline
        \textbf{Category} & \textbf{Num Examples} \\ [0.5ex] 
        \hline
        Shopping & 46  \\ 
        \hline
        Restaurant & 27 \\
        \hline
        Media Sharing & 8  \\ 
        \hline
        News & 55  \\ 
        \hline
        Movies & 14 \\
        \hline
        Transportation & 55  \\ 
        \hline                 
        Real Estate & 20  \\ 
        \hline 
        Hotel & 12 \\
        \hline
        Education \& Information & 20 \\ 
        \hline                 
        Landmarks & 10  \\ 
        \hline                 
        Other & 24  \\ 
        \hline                 
        \textbf{Total} & \textbf{292} \\ 
        \hline
    \end{tabular}
     \caption{Distribution of examples of Trace QA for each category}
     \label{tab:stats_trace}
\end{table}

    \clearpage
    \section{Prompts}
\label{app:prompts}
In this section, we show the prompts used for evaluating the different MLLMs. These prompts correspond to the results presented in Table~\ref{tab:all_results}.

\newenvironment{allintypewriter}{\ttfamily}{\par}

\subsection{Zero-shot prompt}
This prompt is used for evaluating the models by directly predicting the final answer. The prompt is as follows: \\ \\

\begin{allintypewriter}
You are given a sequence of screens and a question. Answer the question using only the information on the screens.
You should directly tell me your answer in the fewest words possible, and do not output any explanation or any other contents. Question: <question>
\end{allintypewriter}

\subsection{Chain-of-thought prompt}
This prompt encourages the models to retrieve the relevant information and then combine it to arrive at the final answer. The prompt is as follows: \\ \\
\begin{allintypewriter}
You are given <num\_screen> screenshots and a question. Your goal is to answer the question according to the screen information only. Please follow the below steps to answer the question.\\
Question: \\
<question>\\

(Screenshots Analysis)\\
First, analyze the contents of each screenshot and list them here.\\

(Question Analysis)\\
What kind of information is needed from each screenshot to answer the question?\\

(Information Extraction)\\
Now, according to the question and analysis, please extract the relevant information from each screen and list them here.\\

(Analyze Information)\\
Based on the question, please analyze how to answer the question using the information above.\\

(Answer)\\
Please generate the answer with the information above. Please exclude any other additional words and information.\\
\end{allintypewriter}

\end{appendices}
\end{document}